\documentclass[twocolumn,prc,showpacs,superscriptaddress,floatfix]{revtex4} 
\usepackage[dvips]{graphicx}
\usepackage{dcolumn}
\usepackage{bm}
\def\Kpi{($K^-,\pi^-$)}
\def\Kpig{($K^-,\pi^- \gamma$)}
\def\piK{($\pi^+,K^+$)}
\def\piKg{($\pi^+,K^+\gamma$)}
\def\von#1{\hbox{$\mbox{{#1}}$}} 
\def\CV{$\check{\von{C}}$erenkov }
\def\lamb#1#2{$^{#1}_{\Lambda}${#2}}
\def\lam#1#2{$^{#1}_{~\Lambda}${#2}}
\def\ggg{$\gamma$}
\def\GE{Ge }
\def\EL#1#2{$^{#1}${#2}}
\def\pp{$p^{-1}_{n}p_\Lambda$}

\begin{document}

\title{Gamma-Ray Spectroscopy of $^{\bf 16}_{\bf ~\Lambda}$O and 
$^{\bf 15}_{\bf ~\Lambda}$N Hypernuclei via the 
$^{\bf 16}$O($\bm{K}^{\bf -}\bm{,}\bm{\pi}^{\bf -}\bm{\gamma}$) reaction}

\author{M.~Ukai}  
\altaffiliation[Present address: ]{Department of Physics, Gifu University,
Gifu 501-1193, Japan.}
\affiliation{Department of Physics, Tohoku University, Sendai 980-8578, Japan} 
\author{S.~Ajimura} 
\affiliation{Department of Physics, Osaka University, Toyonaka 560-0043, Japan}
\author{H.~Akikawa} 
\altaffiliation[Present address: ]{Accelerator Laboratory, KEK, 
Tsukuba 305-0801, Japan.} 
\affiliation{Department of Physics, Kyoto 
University, Kyoto 606-8502, Japan} 
\author{D.~E.~Alburger} 
\affiliation{Brookhaven National Laboratory, NY 11973, USA} 
\author{A.~Banu} 
\affiliation{GSI, Darmstadt D-64291, Germany} 
\author{R.~E.~Chrien} 
\affiliation{Brookhaven National Laboratory, NY 11973, USA} 
\author{G.B.~Franklin} 
\affiliation{Carnegie Mellon University, Pittsburgh, PA 15213, USA} 
\author{J.~Franz}   
\affiliation{Department of Physics, University of Freiburg, Freiburg 79104,  
Germany} 
\author{O.~Hashimoto} 
\affiliation{Department of Physics, Tohoku University, Sendai 980-8578, Japan} 
\author{T.~Hayakawa} 
\affiliation{Department of Physics, Osaka University, Toyonaka 560-0043, Japan}
\author{H.~Hotchi} 
\altaffiliation[Present address: ]{Japan Atomic Energy Agency,
Tokai 319-1195, Japan.} 
\affiliation{Brookhaven National Laboratory, NY 11973, USA} 
\author{K.~Imai} 
\affiliation{Department of Physics, Kyoto University, Kyoto 606-8502, Japan} 
\author{T.~Kishimoto} 
\affiliation{Department of Physics, Osaka University, Toyonaka 560-0043, Japan}
\author{M.~May} 
\affiliation{Brookhaven National Laboratory, NY 11973, USA} 
\author{D.~J.~Millener} 
\affiliation{Brookhaven National Laboratory, NY 11973, USA} 
\author{S.~Minami} 
\altaffiliation[Present address: ]
{GSI, Darmstadt D-64291, Germany.} 
\affiliation{Department of Physics, Osaka University, Toyonaka 560-0043, Japan}
\author{Y.~Miura} 
\affiliation{Department of Physics, Tohoku University, Sendai 980-8578, Japan} 
\author{T.~Miyoshi} 
\altaffiliation[Present address: ]{Institute of Materials Structure Science, 
KEK, Tsukuba 305-0801, Japan.}
\affiliation{Department of Physics, Tohoku University, Sendai 980-8578, Japan} 
\author{K.~Mizunuma}  
\affiliation{Department of Physics, Tohoku University, Sendai 980-8578, Japan} 
\author{T.~Nagae} 
\altaffiliation[Present address: ]
{Department of Physics, Kyoto University, Kyoto 606-8502, Japan.} 
\affiliation{Institute of Particle and Nuclear Studies, KEK, 
Tsukuba 305-0801, Japan} 
\author{S.~N.~Nakamura}  
\affiliation{Department of Physics, Tohoku University, Sendai 980-8578, Japan} 
\author{K.~Nakazawa}  
\affiliation{Department of Physics, Gifu University, Gifu 501-1193, Japan} 
\author{Y.~Okayasu}   
\affiliation{Department of Physics, Tohoku University, Sendai 980-8578, Japan} 
\author{P.~Pile} 
\affiliation{Brookhaven National Laboratory, NY 11973, USA} 
\author{B.~P.~Quinn}  
\affiliation{Carnegie Mellon University, Pittsburgh, PA 15213, USA} 
\author{A.~Rusek} 
\affiliation{Brookhaven National Laboratory, NY 11973, USA} 
\author{Y.~Sato}  
\affiliation{Institute of Particle and Nuclear Studies, KEK, 
Tsukuba 305-0801, Japan} 
\author{R.~Sutter} 
\affiliation{Brookhaven National Laboratory, NY 11973, USA} 
\author{H.~Takahashi} 
\altaffiliation[Present address: ]{Institute of Particle and Nuclear 
Studies, KEK, Tsukuba 305-0801, Japan.}
\affiliation{Department of Physics, Kyoto University, Kyoto 606-8502, Japan} 
\author{L.~Tang}  
\affiliation{Department of Physics, Hampton University, Hampton, VA 23668, USA}
\author{H.~Tamura}  
\affiliation{Department of Physics, Tohoku University, Sendai 980-8578, Japan} 
\author{K.~Tanida} 
\altaffiliation[Present address: ]
{Department of Physics, Kyoto University, Kyoto 606-8502, Japan.} 
\affiliation{RIKEN, Wako 351-0198, Japan}
\author{S.~H.~Zhou} 
\affiliation{Department of Physics, China Institute of Atomic Energy, 
P.~O.~Box 275(80), Beijing 102413, China} 
 
\collaboration{E930('01) collaboration} 
\date{\today}
\begin{abstract}
The bound-state level structures of the $^{16}_{~\Lambda}$O 
and $^{15}_{~\Lambda}$N  
hypernuclei were studied by $\gamma$-ray spectroscopy using a germanium 
detector array (Hyperball) via the $^{16}$O ($K^-, \pi^- \gamma$) reaction. 
A level scheme for $^{16}_{~\Lambda}$O  
was determined from the
observation of three $\gamma$-ray transitions from the doublet of states
($2^-$,$1^-$) at $\sim 6.7$ MeV to the ground-state doublet ($1^-$,$0^-$).
The $^{15}_{~\Lambda}$N   hypernuclei were produced via proton emission from 
unbound states in $^{16}_{~\Lambda}$O . 
Three $\gamma$ -rays were observed and
the lifetime of the $1/2^+;1$ state in $^{15}_{~\Lambda}$N   was measured by
the Doppler shift attenuation method. By comparing the experimental 
results with shell-model calculations, the spin-dependence of the 
$\Lambda N$ interaction is discussed. 
In particular, the measured $^{16}_{~\Lambda}$O ground-state doublet 
spacing of 26.4 $\pm$ 1.6 $\pm$ 0.5 keV determines a small but nonzero 
strength of the $\Lambda N$ tensor interaction.

\end{abstract}

\maketitle
\section{Introduction}
\label{sec:introduction}

 We have performed a series of experiments on the \ggg -ray 
spectroscopy of $\Lambda$ hypernuclei using a germanium detector 
array called Hyperball \cite{tamura00,akikawa02,tamura05}. 
The main purpose of these experiments is 
to study the spin dependence of the $\Lambda N$ interaction via 
precise  measurements of level spacings in $p$-shell hypernuclei.
In this paper, we report on results for the \lam{16}{O} and 
\lam{15}{N} hypernuclei investigated via the 
\EL{16}{O}\Kpi\ reaction at the Brookhaven National Laboratory.

\subsection{$\bm{\Lambda}\bm{N}$ spin-dependent interactions}
\label{sec:intro-calculation}
  
 In principle, interactions between baryons can be studied by  
baryon-baryon scattering experiments. However,  such scattering 
experiments are extremely difficult except for the nucleon-nucleon 
($NN$) case due to the short lifetimes of the other baryons. 
On the other hand, $\Lambda$-hypernuclear data can provide 
information on the $\Lambda N$ interaction.  

 Free hyperon-nucleon ($YN$) interaction models have been 
theoretically constructed as extensions of $NN$ interaction 
models by assuming flavor SU(3) symmetry~\cite{rijken99,rijken06}. 
Effective $YN$ interactions have often been approximated 
as a G-matrix derived from the free $YN$ interaction, either in 
nuclear matter as a function of density \cite{yamamoto94} or using 
the Pauli exclusion operator for finite nuclei \cite{kuo94}. 
Many-body effects entering into the calculation of the effective 
interaction for a finite shell-model space from the G-matrix are 
small, except for the effects of coupling $\Lambda$-hypernuclear 
and $\Sigma$-hypernuclear states ($\Lambda$-$\Sigma$ coupling) 
\cite{tzeng99}. This is in part due to the lack of Pauli 
blocking for a $\Lambda$ in a $\Lambda$ hypernucleus and in part
because the $\Lambda N$ interaction is relatively weak because  
one-pion exchange is forbidden due to isospin conservation.
On the other hand, the $\Lambda N$--$\Sigma N$ interaction
can be mediated by one-pion exchange and it is found to play a 
significant role in the energy-level spacings of s-shell hypernuclei 
despite the $\sim 80$ MeV  $\Lambda$--$\Sigma$ mass difference
\cite{akaishi00,hiyama02,nogga02,nemura02}. 

 A hypernucleus (\lamb{A}{Z}) has a spin-doublet ($J\pm1/2$) 
structure when the core nuclear (\EL{A-1}{Z}) level
has non-zero spin ($J\ne 0$) and the $\Lambda$ is in the 0$s$ 
orbit. In this simple weak-coupling limit, the splitting of the
doublet is due only to $\Lambda N$ interactions that involve
the $\Lambda$ spin (see Eq.~(\ref{lneff}) below) together with 
a contribution from $\Lambda$--$\Sigma$ coupling. Therefore, 
such spin-doublet structures in hypernuclei provide us with 
information on the spin-dependence of the $\Lambda N$ interaction.

The $\Lambda N$ effective interaction \cite{gsd71,millener85}
can be written in the form
\begin{eqnarray}   
V_{\Lambda N} = & & V_0(r) + V_\sigma(r)\,\bm{s}_N\cdot
 \bm{s}_\Lambda \nonumber \\
& & + V_\Lambda(r)\,\bm{l}_{N\Lambda}\cdot \bm{s}_\Lambda
 +  V_{N }(r)\,\bm{l}_{N\Lambda}\cdot \bm{s}_{N} \nonumber \\
& & + V_T(r)\,[3(\bm{\sigma}_N\cdot\hat{\bm{r}}) 
(\bm{\sigma}_\Lambda\cdot \hat{\bm{r}})
- \bm{\sigma}_N\cdot\bm{\sigma}_\Lambda].
\label{lneff}
\end{eqnarray}
The terms correspond to the spin-averaged and spin-spin 
central, the $\Lambda$-spin-dependent spin-orbit ($\Lambda$-spin-orbit),
the nucleon-spin-dependent spin-orbit ($N$-spin-orbit), and the 
tensor interactions, respectively. For $s_\Lambda$ configurations 
in $p$-shell hypernuclei, there are five independent two-body matrix 
elements $\langle p_Ns_\Lambda|V_{\Lambda N}|p_N s_\Lambda\rangle$ that 
can be written in terms of radial integrals associated with each of 
these terms. The radial integrals are conventionally denoted by the 
parameters  $\overline{V}$, $\Delta$, $S_\Lambda$, $S_N$ and $T$, 
respectively~\cite{gsd71,millener85}, with  $S_\Lambda$ and 
$S_N$ the coefficients of $\bm{l}_N\cdot \bm{s}_\Lambda$
and $\bm{l}_N\cdot \bm{s}_N$ (because $\bm{l}_{N\Lambda}$ is proportional 
to $\bm{l}_N$ for a $\Lambda$ in an $s$ orbit).  Since $\overline{V}$ 
contributes equally to all levels, the level spacings are given by 
a linear combination of the four spin-dependent parameters 
and corresponding core level spacings. As noted above, only
$\Delta$, $S_\Lambda$, and $T$ contibute to doublet spacings in the
weak-coupling limit.

 For the same p-shell model space, the $\Lambda N$--$\Sigma N$ 
interaction can be written and parametrized in the same way as
the $\Lambda N$ effective interaction~\cite{millener05}. 
For a fixed $\Lambda$--$\Sigma$ coupling interaction, the four 
$\Lambda N$ parameters that govern the spin dependence of the
interaction can be determined phenomenologically to fit various 
$p$-shell hypernuclear level spacings. In particular, a level 
spacing which is dominantly given by one parameter provides the 
parameter value almost independently of the others. For example, 
if the core state has $L\!=\!0$ only $\Delta$ contributes to
the doublet spacing. This is the case for the ground-state doublet 
spacing in \lamb{7}{Li} ($3/2^+, 1/2^+$) due to an almost 
pure $^3S_1$ configuration for the core \EL{6}{Li}($1^+$). 
Similarly, if the \EL{8}{Be}($2^+$) core state for the
excited-state doublet in \lamb{9}{Be} ($3/2^+, 5/2^+$) has
$L\!=\!2$ and $S\!=\!0$, only $S_\Lambda$ contributes to
the doublet spacing. Then, by comparing the experimentally 
determined parameter values with the predicted values by  
theoretical models of $YN$ interactions, we can test the 
validity of the models.

However, effects of $YN$ interactions on hypernuclear levels are small.
In particular, spin-doublet spacings of $p$-shell and heavier 
hypernuclei are expected to be much smaller than 1 MeV. 
In some cases, hypernuclei have quite small spin-doublet 
spacings of the order of a few tens of keV leading to the
use of the term ``hypernuclear fine structure''.
Therefore, an energy resolution of the order of several keV  is  
necessary for spectroscopic studies  to resolve the level spacings.
Thus, \ggg -ray spectroscopy with germanium (Ge) detectors has 
exclusive access to these structures.

 For this purpose, a Ge detector array
dedicated to the \ggg -ray spectroscopy of hypernuclei, Hyperball, 
was built in 1998 \cite{tamura98} and a project to investigate precise 
structure of hypernuclei started.

\subsection{Previous studies}
\label{sec:intro-previous}

 The energy levels of $\Lambda$ hypernuclei were studied by 
\Kpi\ and \piK\ reaction spectroscopy and by \ggg -ray spectroscopy 
with NaI counters prior to the start of the Hyperball project in 1998.

 Information on the spin-spin interaction can be obtained from the 
observations of the $\sim 1.1$ MeV spin-flip M1 transitions between 
the ground-state doublet ($1^+ \to 0^+$) in \lamb{4}{H} and 
\lamb{4}{He} with NaI counters~\cite{bedjidian79}. However, it has been
shown that $\Lambda$-$\Sigma$ coupling can make a large contribution 
to these doublet spacings~\cite{gibson88}. Recently, four-body
calculations have been performed~\cite{akaishi00,hiyama02,nogga02,nemura02}
for a number of the Nijmegen $YN$ potential models with the result that
both the spin-spin interaction and $\Lambda$-$\Sigma$ coupling make
significant (comparable) contributions to the splitting.
 
 A small $\Lambda$-spin-orbit interaction was reported for the first time
based on the fact that the splitting between the $p_{1/2}$ and $p_{3/2}$ 
substitutional states in \lam{16}{O} is close that of the
underlying hole states of the $^{15}$O core~\cite{bruckner78,bouyssy79}.
Afterwards, the \ggg -ray spectroscopy of \lamb{9}{Be} 
with NaI counters suggested a very small spin-orbit strength
corresponding to $|S_\Lambda |< 0.04$ MeV from a limit
of $< 100$ keV for the spacing of the 3-MeV excited-state doublet 
($3/2^+, 5/2^+$) based on the fact that the width of the peak containing
both $\gamma$ rays was comparable with the resolution of $\sim 160$
keV~\cite{may83}. 

 In 1998, high-precision \ggg -ray spectroscopy experiments for      
\lamb{7}{Li} and \lamb{9}{Be} were carried out with Hyperball 
using the \EL{7}{Li}\piKg\ and the \EL{9}{Be}\Kpig\ reactions, 
respectively~\cite{tamura00,akikawa02}. The ground-state doublet 
spacing of 692 keV in \lamb{7}{Li} provided a spin-spin parameter
value of $\Delta \sim 0.5$ MeV without the inclusion of 
$\Lambda$-$\Sigma$ coupling. In the $p$ shell, the scale of energy shifts
due to $\Lambda$-$\Sigma$ coupling is expected to be roughly a factor
of 4 smaller (with a strong state dependence) than 
that for the $A=4$ hypernuclei~\cite{millener05}, and provides only 
about 12\% of the ground-state doublet spacing in \lamb{7}{Li}. 
A strength of $S_N \sim -0.4$ MeV was also established from the 
excitation energy of the $5/2^+$ state in \lamb{7}{Li}~\cite{tamura00}.  

 In the next experiment, the rather small spacing of the $3/2^+, 5/2^+$
doublet in \lamb{9}{Be} noted above was resolved~\cite{akikawa02}.
The $3/2^+$ state has been determined to be the upper member of the
doublet based on $^{10}$B target data from the present experiment
\cite{tamura05, millener05}. The spacing of 43(5) keV leads to
$-0.02 < S_\Lambda < -0.01$ MeV \cite{tamura05}. The sign
and magnitude of $S_\Lambda$ are consistent with the ordering 
and spacing of the $\Lambda p_{3/2}$ and $\Lambda p_{1/2}$ states 
in \lam{13}{C} measured with a NaI counter array~\cite{kohri02}.

 The energy-level spacings discussed above are not very sensitive
to the tensor interaction. Consequently, for the derivation of other 
three parameters ($\Delta, S_\Lambda$ and $S_N$), $T$ was assumed to be 
in the range of values 0.01 -- 0.06 MeV predicted using
$\Lambda N$ interactions from the Nijmegen one-boson-exchange 
(OBE) models (NSC97f, NSC89, ND and NF) in $G$-matrix 
calculations~\cite{millener85,millener99}. With $T$ taken to
be 0.030 MeV, the remaining  $\Lambda N$ parameters that fit almost 
perfectly the four bound excited states of \lamb{7}{Li}~\cite{ukai06} 
are (parameters in MeV)
\begin{equation}
\Delta= 0.430\quad S_\Lambda =-0.015\quad S_N = -0.390\; .
\label{eq:param7}
\end{equation}

\subsection{The $\bm{\Lambda}\bm{N}$ tensor interaction}
\label{sec:intro-tensor}

 Among the four $\Lambda N$ spin-dependent interactions,
only direct information on the $\Lambda N$ tensor interaction 
has not been obtained experimentally.
The derivation of the tensor interaction strength is important
not only to complete the set of parameters but also to remove 
the theoretical assumption implicit in the
derivation of the other three parameter values.

The strong $NN$ tensor interaction is well understood by 
one-pion exchange, reduced at short distances by rho exchange.
But in the case of $\Lambda N$, these exchanges are forbidden due to 
the zero isospin of the $\Lambda$. The corresponding pseudo-scalar
$K$ and vector $K^*$ exchanges are allowed but cancel strongly
because their masses are more similar than those of the $\pi$ and 
the $\rho$~\cite{millener85}. As a result, the OBE models 
predict small strengths for the tensor interaction. This, and
a relatively small variation for different models, can be seen
from Table~XI of Ref.~\cite{rijken99} for the Nijmegen hard-core
and soft-core models, and from Table~XX of Ref.~\cite{rijken06}
for the extended soft-core models.

 The ground-state doublet spacings in $p_{1/2}$-shell hypernuclei
have large contributions from the tensor interaction. For example,
in the simplified $jj$ coupling model~\cite{gsd71,millener85}, the 
ground-state doublet spacings of \lam{16}{O} and \lam{14}{N}
are given by 
\begin{equation}
\label{p12}
E(1^-)-E(0^-)=-1/3 \Delta+ 4/3  S_\Lambda+8T \ ,
\end{equation}
while the $3/2^+$, $1/2^+$ spacing in \lam{15}{N} is 1.5 times larger. 
In contrast, the spacing of the doublet based on the $p_{3/2}$-hole 
state of \EL{15}{O},
\begin{equation}
E(2^-)-E(1^-)= 2/3 \Delta + 4/3 S_\Lambda 
- 8/5 T \ ,
\label{p32}
\end{equation}
has a coefficient of $T$ relative to $\Delta$ that is an order
of magnitude smaller than for the $p_{1/2}$ doublet in 
Eq.~(\ref{p12}).

 However, the difficulty of determining the tensor interaction 
strength is due not only to its small value but also to the 
small spacings of the spin doublets for which $T$ contributes
strongly. Since $T$ is expected to have a small positive value, and 
$\Delta$ has a relatively large positive value, the contributions of 
$\Delta$ and $T$ are expected to cancel strongly in $p_{1/2}$-shell
hypernuclei, as can be seen from Eq.~(\ref{p12}).
Hence the spacings are expected to be quite small and the ordering of
the states in a doublet can be uncertain. A small spacing also means
that the lifetime of the upper member of a doublet may be long
compared with the lifetime for weak decay ($\sim 200$ ps). 
In this situation, the only safe way to measure a doublet spacing is 
to measure the energies of $\gamma$-rays feeding both members of the 
doublet from a higher level.
 
\subsection{Motivation for the $^{\bf 16}$O($\bm{K}^{-}\bm{,}
\bm{\pi}^{-}\bm{\gamma}$)$^{\bf 16}_{~\bm{\Lambda}}$O 
reaction}
\label{sec:intro-exp}

\begin{figure}
\centerline{\includegraphics[width=8.5cm]{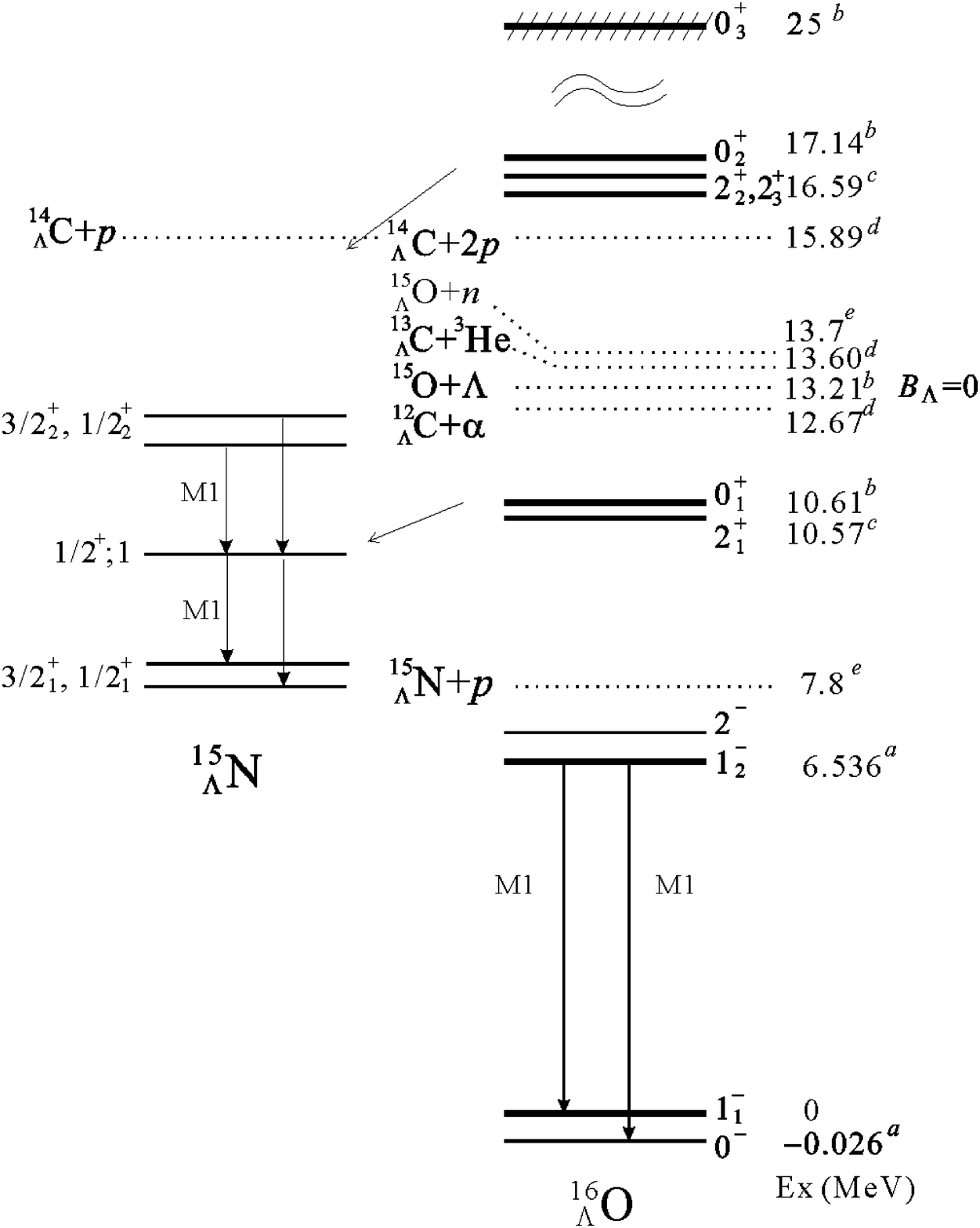}}
\caption{Expected level schemes and \ggg -ray transitions of 
\lam{16}{O} and \lam{15}{N} from the \EL{16}{O}\Kpi\ reaction.
The $1^-$ states and $0^+$ states of \lam{16}{O} (thick lines) 
are most strongly populated by this reaction.
The supscript $a$ shows excitation energies from the present experiment,
$b$ from \cite{bruckner78} and $c$ from \cite{hashimoto98}.
The particle-decay thresholds, except for \lam{15}{N}+p and 
\lam{15}{O}+ n (see text), are determined from emulsion 
data~\cite{emulsion} assuming
$B_\Lambda = 13.21$ MeV for \lam{16}{O}.}
\label{onrelate}
\end{figure}

 In order to study the $\Lambda$N tensor interaction,
the \Kpi\ reaction on $^{16}$O is ideal because
bound states of both \lam{16}{O} and \lam{15}{N} can be produced.
Figure~\ref{onrelate} shows the level schemes of \lam{16}{O} and
\lam{15}{N} (the isospins of 1/2 for \lam{16}{O} and 0 for \lam{15}{N} 
are omitted). Predicted \ggg -ray transitions are also shown.

\begin{figure}
\centerline{\includegraphics[width=7cm]{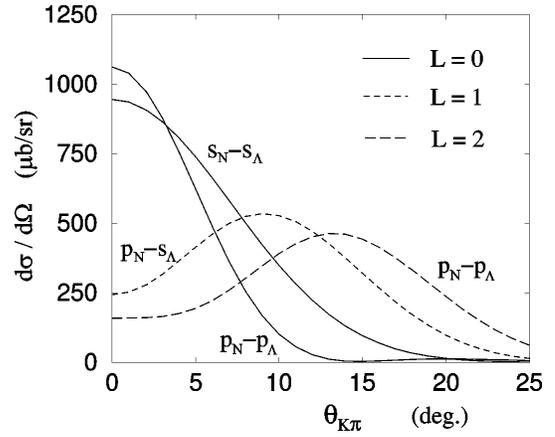}}
\caption{Angular distributions, calculated in distorted-wave 
impulse approximation~\cite{auerbach83}, for the 
\EL{16}{O}$(K^-\pi^-)$\lam{16}{O} reaction at $p_K = 900$ MeV/c as 
a function of the laboratory scattering angle ($\theta_{K\pi}$). 
The cross sections are for the $1^-_2$, $0^+_1$, $2^+_2 + 2^+_3$, 
and $0^+_3$ states shown in Fig.~\ref{onrelate}. For the $1^-_1$, 
$0^+_2$, and $2^+_1$ states, the first three cross sections should 
be multiplied by 0.61, 1.65, and 0.60, respectively.}
\label{dwia}
\end{figure}

 For kaon momenta less than 1~GeV/c, the \Kpi\ reaction proceeds
predominantly by non-spin-flip ($\Delta S\! =\! 0$) transitions.
Theoretical cross sections calculated in distorted-wave impulse
approximation (DWIA) \cite{auerbach83} for all the natural-parity 
states in Fig.~\ref{onrelate} are shown in Fig.~\ref{dwia} for the incident
kaon momentum used in the present experiment ($p_K = 900$ MeV/c).
Substitutional states, in which a neutron is replaced by a $\Lambda$ 
without changing the orbit, are strongly populated at forward angles 
($\theta \leq 5^\circ$) via $\Delta L\! =\! 0$ transitions. The three 
$0^+$ states shown in Fig.~\ref{onrelate} are of this nature (see below)
and were observed, along with the two $1^-$ states ($\Delta L\! =\! 1$), 
at CERN~\cite{bruckner78}. The excitation energies given for the $0^+$
states in Fig.~\ref{onrelate} come from a reanalysis of the CERN 
data~\cite{davis}. The $1^-$ and $2^+$  ($\Delta L\! =\! 2$) states 
are seen strongly in the \EL{16}{O}\piK\lam{16}{O} 
reaction~\cite{hashimoto98, hashtam06} and can be produced in the
\Kpi\  reaction at larger angles (see Fig.~\ref{dwia}).
The excitation energies of the $2^+$ states are taken from 
Ref.~\cite{hashtam06} while the energies of the negative-parity states
come from the present experiment~\cite{ukai04}. The thresholds
in Fig.~\ref{onrelate} depend on the $\Lambda$ binding energies
($B_\Lambda$ values) for the hypernuclei involved and these are
not reliably determined from emulsion data~\cite{emulsion}
(few events with ambiguous interpretations).

 The negative-parity states of \lam{16}{O} in Fig.~\ref{onrelate} 
have dominant $p^{-1}_n s_\Lambda$ configurations while
the positive-parity states have \pp \ configurations except for 
the $0^+_3$ state. The $0^+_1$ and $0^+_2$ states are the 
$p$-substitutional states while the $0^+_3$ state has an
$s^{-1}_n s_\Lambda$ configuration and is the $s$-substitutional 
state. Both $1^-$ states in Fig.~\ref{onrelate} are particle bound.
The 6.6-MeV excited $1^-_2$ state is expected to decay to 
both ground-state doublet members ($1^-_1, 0^-$) by M1 transitions. 
The upper level of the ground-state doublet is expected to decay 
to the lower level by a spin-flip M1 transition. However, for a 
small spacing of less than 100 keV, the detection efficiency of 
Hyperball becomes small. In addition, the weak decay of $\Lambda$ 
in nuclei ($t_{1/2}\sim 200$ ps) would compete with the M1 transition 
for a spacing of less than 100 keV. Thus, we need to detect both 
\ggg \ rays, $1^-_2 \to 1^-_1$ and $1^-_2 \to 0^-$ and determine 
the doublet spacing from the energy difference between these \ggg \ rays.
The spin ordering of the doublet can be determined from the 
branching ratio for these \ggg\ transitions. Since the spin-flip cross 
section is expected to be much smaller than the non-spin-flip one, 
the yield of \ggg \ transitions from the $2^-$ state, which
should decay mainly to the $1^-_1$ state by an $M1$ transition,
is expected to be small. 
 
 It has been pointed out \cite{gal83} that bound states of 
\lam{15}{N} can be produced following proton emission from 
particle-unbound states in \lam{16}{O}. The proton-emission 
threshold of \lam{16}{O} is expected to be at about $E_{x} 
\sim$ 7.8 MeV based on systematics for the difference of the 
$B_\Lambda$ values for \lam{16}{O} and \lam{15}{N}.
Making the same approximation and adding the 5.9 MeV 
between the neutron and proton thresholds in $^{15}$O gives 
$\sim 13.7$ MeV for the neutron-emission threshold in \lam{16}{O}.

The full $1\hbar\omega$ wave functions for the positive-parity 
states of \lam{16}{O} contain admixtures of the following types of
configurations,
\begin{equation}
\label{config}
\alpha | (s^4 p^{11})p_\Lambda> + \beta|(s^4 p^{10} sd)s_\Lambda>
+ \gamma|(s^3 p^{12})s_\Lambda>.
\end{equation}
These states can decay by the $s$-wave and $d$-wave proton 
emission to the positive-parity states in \lam{15}{N} via the
$(s^4 p^{10} sd)s_\Lambda $ components in their wave functions.
The lowest five-positive parity states of \lam{15}{N} are shown 
in Fig.~\ref{onrelate}. For the same reason as in \lam{16}{O}, 
direct observation of the spin-flip M1 transition 
between the ground-state doublet members may not be possible.
Therefore, to observe \ggg \ rays from \lam{15}{N},
the upper doublet ($3/2^+_2, 1/2^+_2$) or the $1/2^+;1$ state
should be produced. The $1^+$ state at $E_{x}=3.95$ MeV in 
\EL{14}{N}, which is the core level for the upper doublet, 
decays by M1 transitions to the $0^+;1$ state at $E_{x}=2.31$ and 
the ground $1^+$ state with the branching rate of 100:4. 
Therefore, the cascade M1 transitions 
$(3/2^+_2, 1/2^+_2) \to 1/2^+;1$ and 
$1/2^+;1 \to (3/2^+_1, 1/2^+_1)$ are likely to be 
observed in \lam{15}{N}.

 The baryonic decay rates of the \pp \ states ($0^+, 2^+$) have been
calculated in the Translationally Invariant Shell Model 
(TISM)~\cite{majling92}. The predictions are that 12\% of the 
$0^+_1$ state decays to the $1/2^+;1$ state and 28\% of the 
$0^+_2$ state to the $1/2^+_2$ state. The $2^+_{2,3}$ states are 
calculated to decay mainly to the $3/2^+_2$ state. 

 Measurement of the ground-state doublet spacings in \lam{16}{O} 
and/or \lam{15}{N} determine a relationship between the values
of $\Delta$ and $T$ from the full shell-model version (which includes
$\Lambda$--$\Sigma$ coupling) of Eq.~(\ref{p12}). A determination
of the energies of more excited states of \lam{16}{O} and \lam{15}{N} 
tests the previously determined spin-dependent interaction strengths.
In particular, the excited-state doublet in \lam{15}{N} is based
on a mainly $^3S_1$ core state, like the $^6$Li ground state,
and the doublet spacing should be large and determined mainly by $\Delta$. 

\section{Experiment}
\label{sec:experiment}

\subsection{Principles}

The \ggg -ray spectroscopy experiment E930('01) on \lam{16}{O} and 
\lam{15}{N} was carried out using the D6 beam line \cite{pile92} at the   
Brookhaven National Laboratory (BNL) Alternating Gradient Synchrotron 
(AGS) for a period of two months in the fall of 2001.

Both \lam{16}{O} and \lam{15}{N} were produced via the 
\EL{16}{O}\Kpi\  reaction with the \ggg\ rays being detected by a large 
acceptance germanium (Ge) detector array called Hyperball.
Incident and outgoing meson momenta were measured
by  magnetic spectrometers and \ggg  -ray spectra were obtained  
by selecting events corresponding to the  \lam{16}{O} mass region.

\subsection{The ($\bm{K}^\mathbf{-}\bm{,}\bm{\pi}^\mathbf{-}$) 
reaction and spectrometers}

\begin{figure*}[bht]
\begin{minipage}[h]{12cm}
\centerline{\includegraphics[width=11cm]{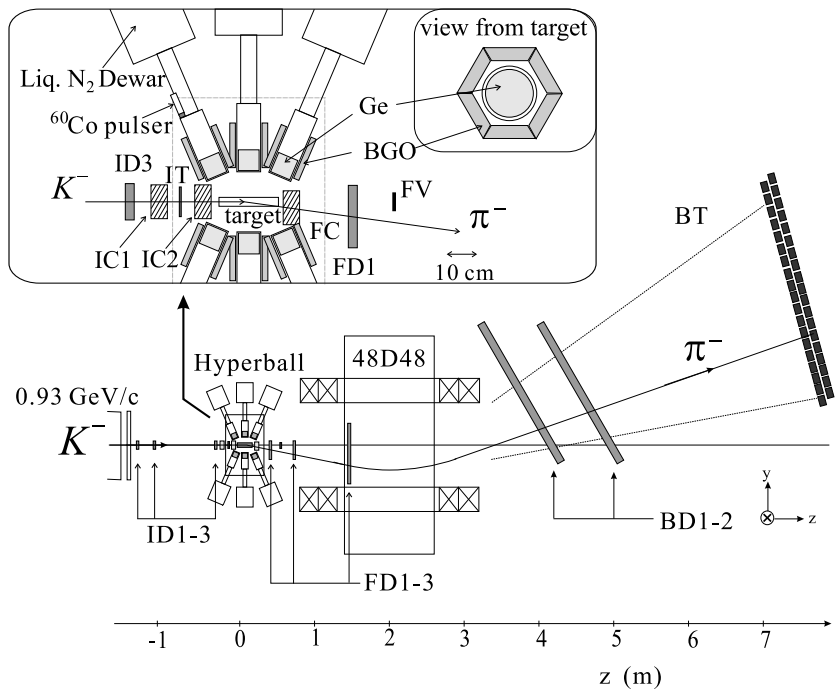}}
\end{minipage}
\begin{minipage}[r]{5.5cm}
\caption{
Schematic view of the experimental setup (side view).
The 48D48 is a dipole magnet, ID's, FD's and BD's are drift 
chambers, IT, FV and BT are plastic scintillation counter hodoscopes
with IT and BT used to measure time-of-flight, and
IC1, IC2 and FC are aerogel \CV counters with $n\! =\! 1.03$.
The Hyperball consisted of fourteen sets of Ge detectors, each surrounded
by six BGO counters. A \EL{60}{Co} pulser was used to monitor the 
Ge detector live time between the beam-on and off periods and 
consisted a 1 kBq \EL{60}{Co} source encapsulated together with
a plastic scintillation counter. 
\ggg -ray events were selected by hits in a Ge detector 
 without hits in the BGO counters. The 20 g/cm$^2$ H$_2$O target  
was irradiated with $4\times10^{10}$ kaons.}
\label{setup}
\end{minipage}
\end{figure*}

A side view of the detection apparatus is shown in  Fig.~\ref{setup}. 
The D6 beam line \cite{pile92}  provided a high-purity and 
high-intensity $K^-$ beam. The incident $K^-$ momentum was set to 
0.93 GeV/c. The energy loss in the beam-line detectors and the target 
medium reduced  the momentum to 0.91 GeV/c at the reaction point.
This $K^-$ momentum was chosen for a balance between hypernuclear 
production yield and Doppler shift effects. The repetition rate of 
the beam spill was  4.6 s and the beam duration time was 1.5 s.
In each spill, $2 \times 10 ^{13}$ 21~GeV/c protons
bombarded the primary platinum target and a $K^-$ beam with 
a typical $K^-/\pi^-$ ratio of 3 was delivered with a typical 
intensity of $2 \times 10^5$. In total, $4 \times 10^{10}$ kaons 
were incident on the 20 g/cm$^2$ water target
(hereafter ``target'' denotes the water target).

 The incident $K^-$ momentum was measured in the downstream part of 
the D6 beam line (not shown in the figure) using the transport matrix
with information from a hit position in the plastic scintillation 
counter hodoscope before the last bending magnet and a straight track 
measured by three drift chambers (ID1--3) after it.
See Ref.~\cite{pile92} for details.  
 

 The spectrometer for scattered particles consisted of a dipole magnet
48D48, which has a pole size of 48''$\times$48'' and a gap 
size of 80 cm. It accepted scattering angles from $-8^\circ$ to $8^\circ$
in the horizontal direction and from $-16^\circ $ to $ 0^\circ $ in the 
vertical direction. The scattered particles were bent vertically (upward).
The momentum was measured by the 48D48 and five drift chambers 
located upstream (FD1--3) and downstream (BD1--2) of the 48D48.
In the present experiment, the 48D48 was operated at 0.8 T, optimized to 
the outgoing $\pi ^-$ momentum of about 0.8 GeV/c.
The field distribution of the 48D48 used in track reconstruction analysis
was calculated by the TOSCA code.


The \Kpi\ reaction events were selected in the trigger level 
by threshold type aerogel \CV counters (AC) with $n\! =\! 1.03$ 
located upstream (IC1 and IC2) and downstream (FC) of the target,
and by time-of-flight in the off-line analysis. Plastic 
scintillation counters MT (located upstream of the last bending magnet 
of the beam line and not drawn in the figure), IT, and BT were 
used to measure the time-of-flight and IT was used as a timing 
reference counter for all the detectors. MT is a horizontally 
segmented hodoscope and BT is a vertically segmented hodoscope. 
The typical flight lengths  between corresponding TOF counters were
15 m (MT -- IT) for incident and 7.9 m (IT -- BT) for scattered particles,
respectively. A plastic scintillation counter (FV), which covered a 
scattering angle less than 3$^\circ$ was located 50 cm downstream from 
the target and used to reject unscattered beam particles and 
very-forward scattered particles in the trigger level.
Events with small reaction angles were also rejected in the off-line analysis
because the vertex resolution is not enough to select reaction events 
in the target. The target was installed between the two AC's
(IC2 and FC) spaced as closely as possible to minimize kaon decay 
events occurring in the target region. Such decay events contributed 
to background, but cannot be eliminated in the off-line analysis 
because their momentum range overlaps with that of hypernuclear 
production. The incident particle identification with 
two stages of AC's (IC1 and IC2) provided an almost pure $K^-$ 
trigger but caused a kaon suppression of 4\%. The efficiency of 
the outgoing particle identification by an AC (FC) for pions was 98\% 
and the misidentification of kaons as pions was 1\%.
However, such kaons were rejected by FV. 

\subsection{Gamma-ray detector, Hyperball}

 Hyperball consisted of fourteen sets of Ge detectors with
bismuth germanate (BGO) scintillation counters. A typical configuration 
and size of the Ge detector and the BGO elements are shown in 
Fig.~\ref{setup}. Each Ge detector was surrounded by six BGO detectors
of 19 mm radial thickness. Each Ge detector had an N-type coaxial crystal of 
about 7 cm $\times$ 7 cm $\phi$ and a relative efficiency to a 
3'' $\times$ 3'' NaI counter of 60\%. The BGO counters were used to 
suppress such background events as Compton scattering, $\pi ^0$ decay, 
and high-energy charged particles. The background from 
$K^- \to \pi^-\pi^0$  decay was particularly serious.
Each Ge detector was equipped with a transistor-reset type preamplifier
and connected to a shaping amplifier with a gated integrator 
(ORTEC 973U \cite{ortec}). This solved the difficulty of operating 
the Ge detectors in a high counting rate and high energy-deposit 
rate environment. We also used timing filter amplifiers 
(ORTEC 579 \cite{ortec}) for timing information.

The end cap of each Ge detector facing the beam was located at 
a distance of 10 cm from the beam axis. The Ge crystals covered a 
total solid angle of 0.25 $\times$  $4\pi$ sr from the target center.
The total photo-peak efficiency in the beam-on period
was measured using \ggg -ray peaks from \ggg -cascade decays
originating in the target and was determined to be               
(1.5 $\pm$ 0.3)\%  for  2.3 MeV \ggg \ rays and 
(4.2 $\pm$ 1.0)\% for 0.7 MeV \ggg \ rays after
including all electronics dead time and all analysis efficiencies.
The efficiency curve as a function of the energy was simulated by
the GEANT code and the absolute scale was adjusted to fit the 
measured efficiencies. The relative efficiency for the beam-on 
and the beam-off periods was measured by the monitoring system 
using triggerable 1 kBq $^{60}$Co sources, each embedded in a 
plastic scintillator connected to a photo-multiplier tube (PMT)
and installed behind a Ge detector as shown 
in Fig.~\ref{setup} (\EL{60}{Co} pulser). These monitoring data 
were taken during both beam-on and beam-off periods with the 
scintillator detecting $e^-$ from $\beta$ decay and the  
corresponding Ge detector detecting \ggg \ rays in coincidence. 
The beam-on/beam-off efficiency ratio for each 
detector was measured to be 90 $\pm$ 5\%.

 Two \ggg -ray sources, \EL{152}{Eu} and  \EL{60}{Co},
were used for energy calibration. However, the \ggg -ray energies of 
available sources were limited to 6.13 MeV, while  the \ggg -ray 
energies from \lam{16}{O} are expected to be around 6.6 MeV.
Therefore, we also used \ggg \ rays from activities (\EL{16}{N}, 
\EL{14}{O}, \EL{24}{Na} and \EL{75}{Ge}$_m$) produced by the beam 
in the target and surrounding materials. In particular, we used 
\ggg -ray peaks following $^{16}$N($\beta ^-$) decay at 6129 keV 
and 7115 keV and their escape peaks. The \EL{16}{N} was most 
likely produced by the \EL{16}{O}($n,p$) reaction on \EL{16}{O} in the
target and surrounding BGO detectors. These \ggg \ rays were observed 
in the Ge-self-trigger data taken during the beam-off period.
Energy calibration was performed in the range from
0.1 to 7.1 MeV. The peak shifts between the beam-on and the 
beam-off periods and the long-term gain shift were also corrected.
The shift between  beam-on/off was typically 1 keV and 
the long-term shift was 2 keV at maximum.

 The response function of the \ggg -ray peak shape after 
summing up the spectra of the fourteen Ge detectors was found 
to be described a Gaussian function up to 6.1 MeV.  
The beam-on energy resolution was  5.7 keV FWHM for 2-MeV 
\ggg  \ rays and 8.6 keV for 6.6-MeV \ggg \ rays.
The energy calibration error was found to be 1.0 keV at 
2 MeV and 1.5 keV for over 5 MeV.

\subsection{Data taking and triggers}

The $K^-$ beam trigger was defined as $\textrm{K}_{in} = \textrm{IT} 
\times\overline{\mbox{IC1}}\times\overline{\mbox{IC2}}$.
The $\pi^-$ scattering trigger was defined as $\textrm{PI}_{out}= 
\textrm{FC} \times \overline{\mbox{FV}} \times\textrm{BT}$.
The data were taken according to the \Kpi\ trigger defined as 
$\textrm{KPI} = \textrm{K}_{in}\times \textrm{PI}_{out}$. The trigger 
rate was typically $1.1 \times 10^3$ per spill. To reduce the trigger 
rate, a second-level trigger was made using information on the Ge 
detector hits. When none of the Ge detectors had hits, data were not 
taken. The second-level trigger rate  was typically $0.3 \times 10^3$ 
per spill.

ADC and TDC data were collected via four FASTBUS crates 
for all the detectors except for the Ge detectors, 
and via FERA bus for Ge detectors, to the corresponding 
VME memory modules (UMEM \cite{umem}) on an event-by-event basis.
The accumulated data were read out by the host computer (LINUX PC)  
in every beam-off period and recorded on a DVD RAM after being 
processed by the host computer. 

Another type of data, triggered only by the Ge detectors 
(Ge-single-trigger data), was also taken in the beam-off period 
of every synchrotron cycle. These data were used for energy 
calibration of the Ge detectors.

\section{Data analysis - ($\bm{K}^\mathbf{-}\bm{,}\bm{\pi}^\mathbf{-}$) 
reaction}
\label{sec:analysis-kpi}

\subsection{Hypernuclear masses}

 The \lam{16}{O} mass ($M_{^{16}_\Lambda{\rm{O}}}$) 
was reconstructed as a missing mass in the \EL{16}{O}\Kpi\  reaction.
The binding energy of a $\Lambda$ in the hypernucleus is defined by  
\begin{equation}
\label{blam}
B_\Lambda = M_{{^{15}\rm{O}}}+M_\Lambda-M_{^{16}_{~\Lambda}\rm{O}},
\end{equation}
where $M_{{^{15}\rm{O}}}$ and $M_\Lambda$ are the masses of the 
core nucleus  (\EL{15}{O}) in its ground state and the $\Lambda$, 
respectively.

The absolute mass scale was calibrated using the $\pi^0$ mass 
reconstructed from the $K^- \to \pi^- \pi^0$ decay. The energy losses 
of the $K^-$ and the $\pi^-$ in the target medium were corrected event 
by event. In the present experiment, the mass resolution was 15 MeV 
(FWHM). We selected events with a reaction angle larger than $2^\circ$
and a reaction vertex point in the target region.

\begin{figure}
\centerline{\includegraphics[width=5.5cm]{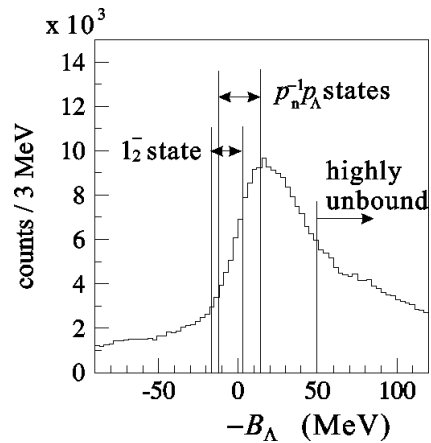}}
\caption{Missing-mass spectrum for the \EL{16}{O}\Kpi\ reaction 
plotted against the  $\Lambda$ binding energy ($B_\Lambda$) for 
those events accompanying  \ggg \ rays with energies of 1.5 -- 7.0 MeV.
The indicated regions show the definitions of 
the $1^-_2$ state region ($-17 < -B_\Lambda < 3$ MeV),
the \pp  \ states region ($-12 < -B_\Lambda < 14$ MeV) and the
highly unbound region ($-B_\Lambda > 50$ MeV). 
\label{oxblam}} 
\end{figure}

Figure \ref{oxblam} shows the missing-mass spectrum from the 
\EL{16}{O}\Kpi\ reaction plotted against the $\Lambda$ binding energy 
($B_\Lambda$) for those events accompanying \ggg  \ rays with energies
in the range from 1.5 MeV to 7.0 MeV. As shown in Fig.~\ref{onrelate}, 
four narrow states ($1^-_1, 1^-_2, 0^+_1$ and  $0^+_2$) and one broad 
state ($0^+_3$), corresponding to the $s^{-1}_N s_\Lambda$ substitutional 
state at around $-B_\Lambda= 12$ MeV \cite{bruckner78}, are expected to 
be produced. However, they can not be resolved in the missing-mass 
spectrum due to the limited mass resolution.

 To observe the \lam{16}{O} \ggg  \ rays, the mass region corresponding 
to the $1^-_2$ state was selected.
The mass of the $1^-_2$ state is near $B_\Lambda = 7$ MeV
from previous experiments \cite{bruckner78,hashimoto98,tamura94}.
Consequently, the $1^-_2$ state region was defined by $-17 < -B_\Lambda 
< 3$ MeV. The 20-MeV gate width corresponds to 88\% of all the $1^-_2$ 
state events for the 15 MeV resolution.
On the other hand, to observe the \lam{15}{N} \ggg  \ rays,
a mass region including the $0^+$ and $2^+$ states (\pp -state region)
was selected. The $0^+_1$ and $0^+_2$ states are at
$B_\Lambda = 2$  and $-4 $ MeV \cite{bruckner78}, as are
the $2^+_1$ and $2^+_{2,3}$ states~\cite{hashimoto98}. 
Therefore, we defined this region to be $-12 < -B_\Lambda < 14 $ MeV 
and this gate width covers 94\% of the events for the
 $0^+_1$ and $0^+_2$ states.
We also defined the highly unbound region as  $-B_\Lambda > 50$ MeV.
These defined regions are shown in Fig.~\ref{oxblam}. 

\subsection{Kinematical conditions}

\begin{figure}[h]
\centerline{\includegraphics[width=6cm]{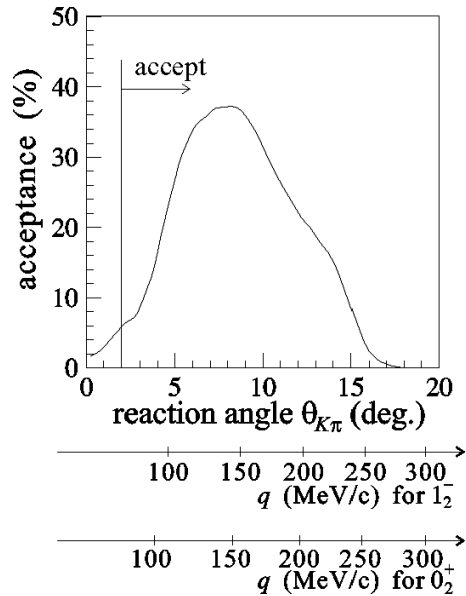}}
\caption{Simulated reaction-angle ($\theta_{K\pi}$)
dependent acceptance of the scattered-particle spectrometer.
Differences in acceptance for the $1^-_2$ ($E_x = 6.6$ MeV) 
and $0^+_2$ ($E_x = 17.1$ MeV) states of \lam{16}{O} are negligibly 
small. Momentum transfers ($q$ MeV/c) for these states corresponding to the 
reaction angles are also shown. Events with reaction angles less than 
2$^\circ$ were rejected in the off-line analysis.}
\label{acceptance}
\end{figure}

Figure~\ref{acceptance} shows the acceptance, obtained by a 
Monte-Carlo simulation, of the scattered-particle spectrometer
as a function of the reaction angle ($\theta_{K\pi}$). The 
momentum transfer [$q$ (MeV/c)] corresponding to $\theta_{K\pi}$
is also shown. We simulated the cases of producing the $E_{x}=6.6$ 
MeV ($1^-_2$) and  $17.1$ MeV ($0^+_2$) states. Differences in the 
$q$-dependent acceptances for these states were found to be 
negligibly small. The $K^-$ momentum at the reaction point was 
distributed around 0.91 GeV/c with a width of 0.06 GeV (FWHM) on 
account of the original beam momentum spread and energy loss 
in the thick target.

 As is evident from Fig.~\ref{dwia}, the cross sections for the 
$0^+$, $1^-$, and $2^+$ states formed via the \EL{16}{O}\Kpi \lam{16}{O} 
reaction at $p_{K^-} = 900$ MeV/c peak at $0^\circ$, $9^\circ$, 
and $13^\circ$, respectively.
Taking into account the angular distribution of these reactions and the
spectrometer acceptance, the initial recoil momenta of \lam{16}{O} 
for the $1^-_2$ and  $0^+_2$ states  were estimated to be  
in the range of $100-250$ MeV/c and $100-200$ MeV/c, respectively.
The recoil momentum of \lam{15}{N} was also calculated to be 
in the range of  $0-250$ MeV/c assuming that the proton was emitted 
isotropically. The corresponding stopping times of these hypernuclei 
in the target medium were calculated using the SRIM code~\cite{srim}
to be in the range of 1.5 -- 2.5 ps and 0 -- 2 ps, respectively.

\section{Data Analysis - gamma rays}
\label{sec:analysis-gamma}

\subsection{Event selection for gamma rays}

Since the timing of a Ge detector hit with respect to the beam 
timing (IT hit) varies with \ggg -ray energy, the timing gate 
width was set as a function of the energy. The gate width was  
50 ns for 0.5 MeV, 30 ns for 2 MeV, and 20 ns for over 5 MeV.
The gate width for the BGO counters was set to be 50 ns so as 
to achieve efficient background suppression without
oversuppression from accidental coincidences.

\begin{figure}[h]
\centerline{\includegraphics[width=6.cm]{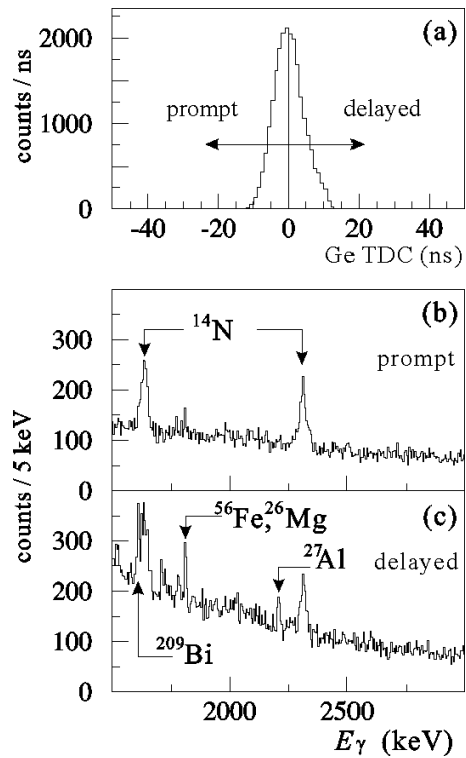}}
\caption{(a) TDC spectrum for a typical Ge detector plotted 
for \ggg -ray energies over 1.5 MeV.
(b) and (c) are the \ggg -ray energy spectra plotted for 
the highly-unbound region in \lam{16}{O} with the prompt 
and delayed timing cuts shown in (a), respectively.}
\label{tdccut}
\end{figure}

To observe and identify statistically weak \ggg -ray peaks,
the background level should be minimized. When gating  the 
particle-unbound region, we found a lot of \ggg \ rays, 
such as \EL{27}{Al} and \EL{56}{Fe} \ggg \ rays around the
2 MeV region, originating from ($n,n'$) reactions in the 
detector and surrounding materials. Since the neutrons were 
assumed to reach these materials a few ns after the triggered 
timing, we made another TDC cut which was used only to 
identify these \ggg \ rays. Figure~\ref{tdccut} (a) shows a 
TDC spectrum for a typical Ge detector plotted for \ggg -ray 
energies over 1.5 MeV. The timing cut conditions, prompt and 
delayed, were defined as shown in Fig.~\ref{tdccut} (a).
Figures \ref{tdccut} (b) and (c) show \ggg -ray spectra 
plotted for the highly-unbound region in \lam{16}{O} with
the prompt and  delayed timing cuts, respectively.
\ggg \ rays from the target materials are 
enhanced in the prompt spectrum (b) and 
\ggg \ rays from the detector materials are enhanced 
in the delayed spectrum (c).

\subsection{Gamma-ray Doppler shifts}

The energies of the expected \ggg \ rays  from \lam{16}{O} and 
\lam{15}{N} are larger than 1 MeV except for the spin-flip 
transitions within the spin doublets. They are all M1 transitions
so that their lifetimes are expected to be much shorter than, or
of the same order as, the stopping time of the recoiling hypernucleus.
Therefore,  each \ggg -ray peak  has a Doppler-broadened shape 
determined by the lifetime of the state, the stopping time, 
the response function of the \ggg \ ray, and 
the kinematical conditions of the experiment.

\subsubsection{Doppler-shift correction}
\label{shift-correction}

 When the lifetime of the initial state is much shorter than 
the stopping time, $\tau \alt 0.1$ ps, the \ggg -ray peak shape is 
fully broadened by the Doppler-shift effect.
If we know the velocity of the recoiling hypernucleus ($\beta$)
and the angle between the hypernuclear velocity 
and \ggg -ray emission ($\theta$) in the laboratory frame,
the shifted \ggg -ray energy can be corrected for by the 
relativistic formula,
\begin{equation}
E^C_\gamma = E^M_\gamma \cdot 
(1-\beta \cos \theta)/\sqrt{1-\beta ^2},
\end{equation} 
where $E^C_\gamma$ and $E^M_\gamma$ are the Doppler-shift corrected
energy and the measured energy, respectively.
The direction of  \ggg  -ray emission in the laboratory frame
was calculated from the 
reaction vertex point and the position of the Ge crystal with a hit.
The velocity ($\beta$) and direction of the hypernucleus
were calculated from the kaon and pion momentum vectors
measured by the incident and outgoing particle spectrometers.

\begin{figure}[thb]
\centerline{\includegraphics[width=7cm]{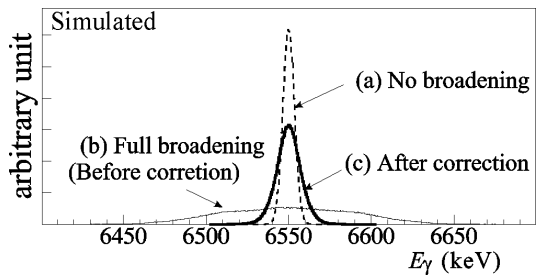}}
\caption{
Simulated  peak shapes of \ggg \ rays
emitted from recoiling \lam{16}{O}, produced 
by the \EL{16}{O} \Kpi reaction,
assuming a \ggg -ray energy of 6.55 MeV.
The figure shows  
(a) not broadened (original response function),
(b) fully Doppler-broadened (before correction) and 
(c) Doppler-shift-corrected peak shapes.
}
\label{ogam}
\end{figure}  

First, we consider the Doppler-shift correction for \lam{16}{O} \ggg \  
rays. Since the \ggg -ray energies of $M1$ transitions from the $1^-_2$ 
state are larger than 6 MeV, the lifetime of the state is expected
to be very short because the lifetime of the corresponding core
level ($3/2^-$) is very short ($< 2.5$ fs).
Figure~\ref{ogam} shows the simulated \ggg -ray peak shapes of \lam{16}{O}
assuming an original energy of 6.55 MeV,
giving (a) the original response function, 
(b) the fully Doppler-broadened shape and 
(c) the shape after Doppler-shift correction for (b).

\begin{figure}[thb]
\centerline{\includegraphics[width=7cm]{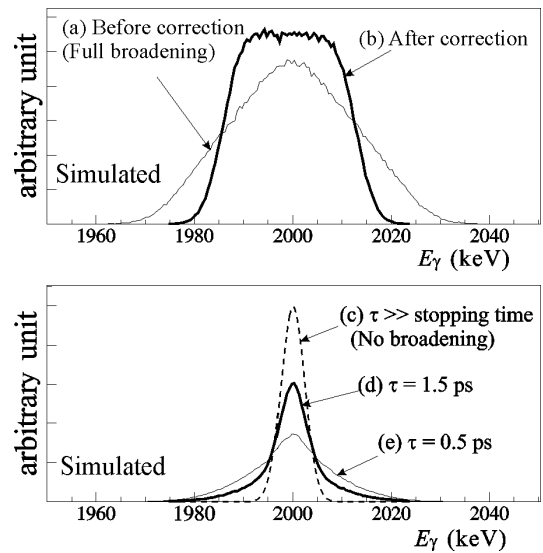}}
\caption{Simulated peak shapes for 2-MeV $\gamma$ rays from 
\lam{15}{N} following proton emission from the 
\lam{16}{O}($0^+_2$) state produced via  the \EL{16}{O} \Kpi reaction.
The upper figure shows 
(a) fully Doppler-broadened and 
(b) Doppler-shift-corrected peak shapes.
The lower figure   
shows the lifetime dependence of the peak shape,
(a) $\tau \gg $ stopping time (no broadening, original response function),
(b) $\tau = 1.5$ ps and  
(c) $\tau = 0.5$ ps. 
}
\label{ngamsim}
\end{figure}

Second, we consider the Doppler-shift correction for \lam{15}{N} 
\ggg \ rays. The lifetimes of the upper-doublet states 
($3/2^+_2, 1/2^+_2$) in \lam{15}{N} are expected to be a few fs 
\cite{millener05}. 
The recoil velocity and the recoil direction of \lam{15}{N} are
changed from those of \lam{16}{O} due to the proton emission.
However, since the mass of the secondary nucleus (\lam{15}{N}) is  
not so different from the mass of primary nucleus (\lam{16}{O}) 
and the emitted proton energy is small, the Doppler-shift correction 
can be also applied to the the secondary nucleus. We  simulated a 
Doppler-shift corrected \ggg -ray peak shape for \lam{15}{N} using 
a simulated $\beta$ for \lam{16}{O}($0^+_2$), assumed to be produced 
according to the calculated angular distribution (see Fig.~\ref{dwia}).
Proton emission was assumed to be isotropic.
We found that the Doppler-shift correction makes a broadened
peak narrower and square-shaped. Figure \ref{ngamsim} shows the 
simulated \ggg -ray peak shapes assuming that a 2-MeV \ggg \ ray  
is emitted from \lam{15}{N} following proton emission;
(a) fully Doppler-broadened and 
(b) Doppler-shift-corrected peak shapes.

 In addition, we also simulated other decay processes such as
\lam{13}{C}+\EL{3}{He} and \lam{12}{C}+$\alpha$.
In these cases, it was found that the correction does not make the
Doppler-broadened peak shapes narrower.

In these simulations, we assumed that hypernuclei emit \ggg \ rays 
isotropically  in the center-of-mass frame. Although angular 
correlations are expected to exist between the scattered $\pi ^-$ 
and \ggg \ rays \cite{dalitz78}, the effect on the corrected energy 
is simulated to be negligibly small.

\subsubsection{Doppler-shift attenuation method}

 Information on the lifetime of a state can be obtained by analyzing 
the partly Doppler-broadened peak shape. This is called 
the Doppler-shift attenuation method (DSAM). In particular,
when the stopping time and the lifetime are of the same order of 
magnitude, DSAM can be used to determine the lifetime.
This method was first applied for hypernuclei to
determine the $B(E2)$ value of the $5/2^+$ excited state of
\lamb{7}{Li}~\cite{tanida01}.

 DSAM can also be  applied to the secondary hypernuclei if the 
recoil momentum is known. Among the states to be produced, the 
\lam{15}{N}($1/2^+;1$) state is expected to have a lifetime of 
the same order as the stopping time ($\sim 2$ ps), an estimate 
for the lifetime of the state being 0.5 ps \cite{millener05}.
However, the absolute mass of \lam{15}{N} is not well determined
and the \lam{15}{N} production rates from the $0^+_1$, $0^+_2$ and 
$s$-substitutional states of \lam{16}{O} cannot be obtained because 
of the limited mass resolution. These ambiguities are
included  in  the systematic errors. The ambiguities  mainly  stem from 
the Q value of the \lam{16}{O} $\to$ \lam{15}{N} + $p$  decay.
Details are given later.

In the simulation, we assumed that \lam{16}{O} decays to \lam{15}{N} 
and a proton isotropically in the center-of-mass frame.
The simulated \ggg -ray peak shape is shown in 
Fig.~\ref{ngamsim} (bottom).
The figure shows the lifetime dependent peak shape for
(a) $\tau \gg$  stopping time (no broadening, original response function),
(b) $\tau =1.5$ ps and (c) $\tau =0.5$ ps.

\section{Results}
\label{sec:results}

\begin{figure*}
\centerline{\includegraphics[width=14cm]{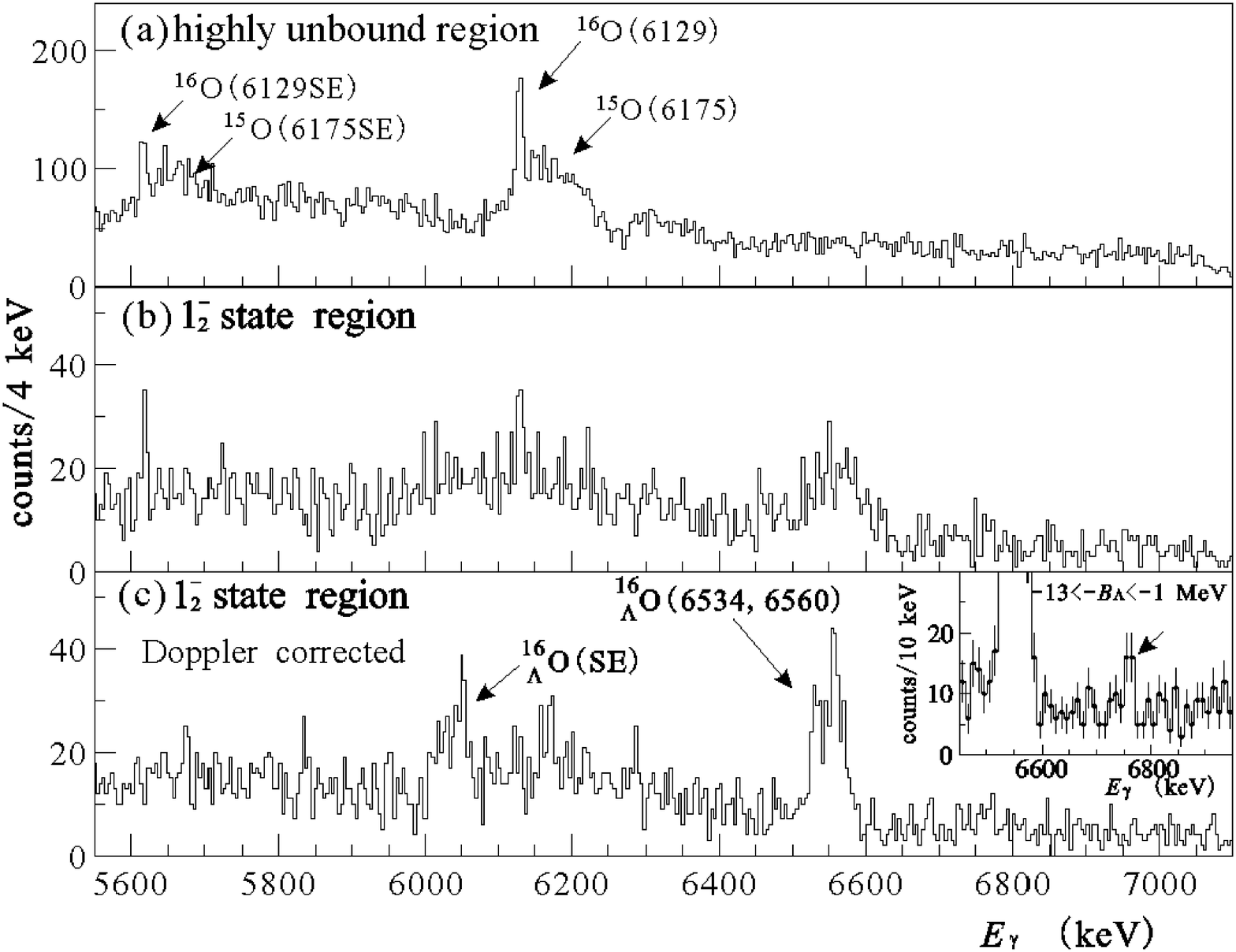}}
\caption{Mass-gated \ggg -ray spectra measured in the \EL{16}{O}\Kpi\ 
reaction at $\theta_{K\pi} > 2^\circ$ for 
(a) the highly-unbound region ($-B_\Lambda > 50$ MeV) and 
(b) the $1^-_2$ state region ($-17 < -B_\Lambda < 3$ MeV).
(c) is the same spectrum as (b) but with event-by-event Doppler-shift
corrections applied. A structure around 6.5 -- 6.6 MeV in (b) 
becomes two narrow peaks after Doppler-shift correction.
The two peaks are assigned as $M1(1^-_2 \to 1^-_1, 0^-)$ transitions.
Single-escape (SE) peaks from these $M1$ transitions can also be seen
in (c), and likewise SE \ggg-ray peaks from \EL{15}{O}(6175 keV) and 
\EL{16}{O}(6129 keV) in (a). Inset (c) shows the spectrum for a narrower 
mass cut ($-13 < -B_\Lambda < -1 $ MeV) which is set to maximize the
signal to noise ratio for the 6534- and 6560-keV $\gamma$-ray peaks.
A gathering of events with a statistical significance  
$3\sigma$ from background appears at 6758~keV.}
\label{oxgam}
\end{figure*}

Some results for the \lam{16}{O} \ggg \ rays have already been 
reported in Ref.~\cite{ukai04} which was focused on
the determination of the ground-state doublet spacing in \lam{16}{O}. 
In this paper, details of the results on the \lam{16}{O} \ggg \ rays 
and additional results on the \lam{15}{N} \ggg \ rays are discussed.

\subsection{Gamma rays from $^{\bf 16}_{~\bm{\Lambda}}$O}
\label{sec:results-16o}

\subsubsection{Gamma-ray spectrum}
\label{gamma16}

Figure~\ref{oxgam} shows the mass-gated $\gamma$-ray energy spectra 
around 6 MeV in coincidence with the \EL{16}{O}\Kpi\ reaction at 
$\theta_{K\pi} > 2^\circ$; 
(a) is plotted for the highly-unbound region ($-B_\Lambda > 50 $ MeV)
without Doppler-shift correction ($E_\gamma^M$),
(b) is for the $1^-_2$-state region ($ -17 < -B_\Lambda < 3 $ MeV)
without Doppler-shift correction  ($E_\gamma^M$), and (c) is for 
the $1^-_2$ state region with Doppler-shift correction ($E_\gamma^C$).
Here, the Doppler-shift correction was applied assuming that 
\ggg \ rays were emitted before the recoiling hypernucleus slowed down.
The highly unbound and the $1^-_2$ state regions are defined in 
Fig.~\ref{oxblam}.

We found \EL{16}{O}(6129 keV) and \EL{15}{O}(6175 keV) \ggg -ray peaks 
in the spectra for both the highly-unbound region (a) and the 
$1^-_2$-state region (b). On the other hand, a bump around 6.6 MeV was 
found only in the spectrum for the $1^-_2$-state region (b)
and it becomes two narrow peaks after the Doppler-shift correction,
 as shown in (c). The peaks are attributed to $M1$ transitions from 
the $1^-_2$ state to both ground-state doublet members ($1^-_1, 0^-$) 
in \lam{16}{O} because the $1^-_2$ state is the only bound state,
other than the $1^-$ member of the ground-state doublet, that 
can be strongly populated by the reaction and \ggg -ray peaks from
other (hyper)nuclei cannot become so narrow after Doppler-shift 
correction (Fig.~\ref{ngamsim}). This argument is supported by
the fact that the shape of the bump before Doppler-shift correction 
in Fig.~\ref{oxgam} (b) is consistent with the fully Doppler-broadened 
peak shape [Fig.~\ref{ogam} (b)]. In addition, the structure observed 
at $\sim 6.04$ MeV in Fig.~\ref{oxgam} (c) has a shape similar to 
the \lam{16}{O} peaks at $\sim$ 6.55 MeV and corresponds to their 
single-escape (SE) peaks.

\begin{figure}[thb]
\centerline{\includegraphics[width=4.86cm]{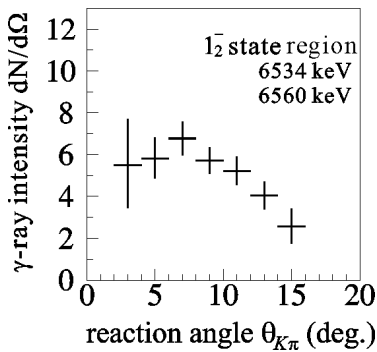}}
\caption{Reaction angle ($\theta_{K\pi}$) dependence of the 
\lam{16}{O} $\gamma$-ray intensities (sum of the 6534- and 6560-keV 
\ggg \ rays) from the $1^-_2$ state in \lam{16}{O} measured in 
the \EL{16}{O}\Kpi\ reaction for the $1^-_2$ state region  
($-17 < -B_\Lambda < 3$ MeV) defined in Fig~\ref{oxblam}.
Shown are the \ggg -ray counts divided by solid angle taking 
into account the spectrometer acceptance (Fig.~\ref{acceptance}).}
\label{oxangdis}
\end{figure}

 We made a fit to the Doppler-shift corrected spectrum with 
the simulated peak shape [Fig.~\ref{ogam} (c)] assuming that the 
background around a full-energy peak has two components; one is 
constant for the whole region below and above the \ggg -ray energy
($E_0$), and the other is caused by  multiple Compton scattering and 
exists only on the lower-energy side with the form 
$a+bE$ for $E <E_0$ and 0 for $E>E_0$. In a simulation 
taking into account the Doppler-shift correction, the step-function 
was smoothed around $E=E_0$ and then used in the fitting.
We obtained for the lower peak an energy  of 
$6533.9 \pm 1.2$(stat) $\pm 1.7$(syst) keV and
a yield of $127\pm15\pm5 $ counts, and for the upper peak, 
$6560.3\pm1.1\pm1.7$ keV and $183\pm16\pm5$ counts.
The systematic errors include the calibration error
and the accuracy of the  Doppler-shift correction.
The excitation energy of the $1^-_2$ state becomes 6561.7 keV after 
applying the nuclear recoil correction to the \ggg -ray energy.
The  energy difference between the two \ggg \ rays, $26.4\pm1.6\pm0.5$ keV,
corresponds to the ground-state doublet spacing.
The relative intensities of the \ggg \ rays are
\begin{equation}
\frac{I_\gamma(6534)}{I_\gamma(6560)} = 0.69 \pm 0.11 \pm 0.10,
\label{o16ratio}
\end{equation}
where the \ggg -ray efficiencies were assumed to be the same. 

 We also found a gathering of events at 6760 keV in the 
$1^-$ state region spectrum (c). To reduce the background, we set a 
narrower mass gate ($-13 < -B_\Lambda < -1$ MeV) to maximize the 
signal to noise ratio, $S^2/N$, for the 6534- and 6560-keV $\gamma$-ray
peaks. Then, we found a peak with a 
statistical significance of $3\sigma$ ($21.0 ^{+7.2}_{-6.5}$ counts) 
at an energy  of $6758 \pm 4 \pm 4$ keV, as shown  
in Fig.~\ref{oxgam} inset (c). The peak width was found to be consistent 
with the simulated Doppler-shift-corrected peak shape.
Applying  the same mass gate, the sum of the $1^-_2 \to 1^-_1, 0^-$ 
\ggg -ray yields was was found to be 262 $\pm$ 24 counts.
The \ggg -ray intensity ratio was 
\begin{equation}
\frac{I_\gamma(6758)}{
I_\gamma(6534)+I_\gamma(6560)}= 0.08 \pm 0.03.
\label{ds01}
\end{equation}

\subsubsection{Angular distributions}
\label{angdist16}

 Figure \ref{oxangdis} shows the reaction-angle ($\theta_{K\pi}$) 
dependence for the sum of the 6534- and 6560-keV \ggg \ rays
observed from the $1^-$ state region, $-17< -B_\Lambda < 3$ MeV, 
defined in Fig.~\ref{oxblam}. The sum of these \ggg-ray yields 
were obtained from a fit using the simulated peak shape with 
the two \ggg -ray energies and the ratio of the two \ggg -ray yields
fixed to be 0.69 from Eq.~(\ref{o16ratio}). The \ggg -ray intensity 
was obtained from the summed \ggg -ray yield divided by a solid angle 
taking into account the angular dependence of the spectrometer 
acceptance (Fig.~\ref{acceptance}). The observed angular distribution 
agrees well with calculated distribution in Fig.~\ref{dwia} for a 
$\Delta L =1$ transition and is thus consistent with the the state 
that emits the 6534- and 6560-keV \ggg \ rays being the $1^-_2$ 
state.

\subsubsection{Level assignments}
\label{assign16}

  The branching ratio $I_\gamma(1^- \to 1^-)/I_\gamma(1^- \to 0^-)$ 
is 0.5 in the weak-coupling limit and 0.41 when the level mixing
is taken into account~\cite{millener05}. The measured relative intensity 
of the 6534- and 6560-keV \ggg \ rays [Eq.~(\ref{o16ratio})] should 
therefore give information on the spin ordering of the doublet members. 
First, a correction has to be made for the fact that the  effective 
\ggg -ray efficiency of the Hyperball is different for the two 
transitions due to angular distribution effects. For pure M1
transitions and a forward reaction angle ($\theta_{K\pi}\sim 0^\circ$) 
\cite{dalitz78},
\begin{equation}
W(\theta_{\pi\gamma}) \propto 
\cases{
(1+\cos ^2 \theta_{\pi\gamma} ) & for  $(1^- \to1^-)$ \cr
(1-\cos ^2 \theta_{\pi\gamma} ) & for  $(1^-\to 0^-)$ \cr 
},
\end{equation}
where $\theta_{\pi\gamma}$ is the angle between the $\pi^-$ and 
the $\gamma$ ray. The \GE detectors were not arranged 
isotropically with respect to the $\pi^-$ direction and
effective efficiencies were calculated from a simulation
with the result $\varepsilon (1^-\!\to\!1^-)/\varepsilon
(1^-\!\to\!0^-)\! =\! 0.80\pm0.05$(syst.). The error comes from  
ambiguities in the real size and the shape of the \GE detectors.
Since we measured the hypernuclear production events at finite
angles, the actual ratio is larger than this estimation. Taking
into account the $\pi\gamma$ correlation, the expected branching
ratios of 0.41 or 0.50 should be corrected to 0.33 or 0.40.
Thus, the measured ratio of  $0.69\pm 0.11\pm 0.10$ is slightly larger than 
expected but still favors a $0^-$ assignment for the lower level 
of the ground-state doublet.

The intensity of the 6758 keV \ggg \ ray is much weaker than 
those of the transitions from the $1^-_2$ state [Eq.~\ref{ds01}].
With the caveat that $2^+$ hypernuclear states based on the 
6.793-MeV $3/2^+$ or 6.859-MeV $5/2^+$ levels of $^{15}$O could 
possibly be excited through small $p^{-1}p_\Lambda$ components in 
their wave functions [cf. Eq.~(\ref{config})], this \ggg -ray 
transition is tentatively attributed to the M1($2^- \to 1^-_1$) 
transition from the $2^-$ member of the $p_{3/2}^{-1}s_\Lambda$ 
doublet, giving an excitation energy of $6786 \pm 4 \pm 4$ keV for 
the $2^-$ state. If this assignment is correct, this is the first 
experimental data to identify directly produced spin-flip and 
non-spin-flip states in $\Lambda$ hypernuclei.  

\subsubsection{Missing-mass spectrum}
\label{mass16}

\begin{figure}
\centerline{\includegraphics[width=8.5cm]{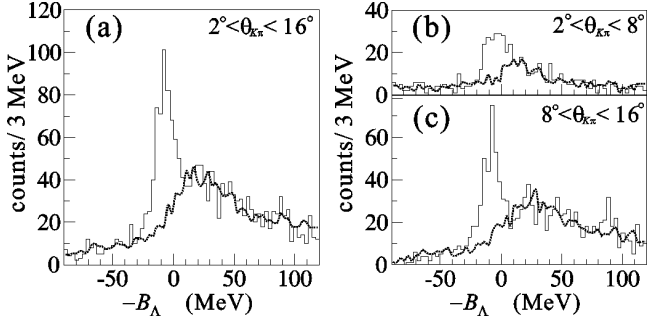}}
\caption{Missing-mass spectra measured in the \EL{16}{O}\Kpi\ reaction
plotted against the $\Lambda$ binding energy ($B_\Lambda$)
for events accompanying a \ggg \ ray with $E_\gamma^C=6500$ -- 6600 keV
for (a) $2^\circ< \theta_{K\pi} < 16^\circ$,
(b) $2^\circ< \theta_{K\pi} < 8^\circ$,  and
(c) $8^\circ< \theta_{K\pi} < 16^\circ$.
The dotted lines show the expected background shapes plotted for events 
accompanying \ggg -rays with $E_\gamma^C=$ 6600 -- 7000 keV and scaled 
to fit the background region ($-90 < -B_\Lambda < -40$ MeV).}
\label{blam1}
\end{figure}

 Figure~\ref{blam1} shows missing-mass spectra as a function
of $\Lambda$ binding energy ($B_\Lambda$) for events accompanying a 
\ggg \ ray with a Doppler-shift corrected \ggg -ray energy 
$E_\gamma^C=6500$ -- 6600 keV corresponding to the 
$1^-_2 \to 1^-_1, 0^-$ transitions. In Fig.~\ref{oxblam} (a), a
prominent peak is observed corresponding to the experimental value of 
$-B_\Lambda = -7$ MeV for the $1^-_2$ state.
By fitting this peak in with a Gaussian, the mass resolution
was determined to be 15 $\pm$ 1 MeV. The mass resolutions 
obtained by fitting the mass spectra in Fig.~\ref{oxblam} (b) and (c)
were $15 \pm 2$ MeV.

\subsubsection{Lifetime analysis}
\label{lifetime16}

The fully broadened \ggg -ray peak shape contains information on 
the lifetime. We fitted the Doppler-shift uncorrected spectrum 
with simulated peak shapes for various lifetimes using
the measured \ggg -ray energies and yield ratio.
Then, we obtained an upper limit for the lifetime of the 
$1^-_2$ state in \lam{16}{O} of $\tau < 0.3$ ps 
at the 68\% confidence level.

\subsection{Gamma rays from $^{\bf 15}_{~\bm{\Lambda}}$N}
\label{sec:results-15n}

\subsubsection{Gamma-ray spectrum}
\label{sec:gamma15spec}

\begin{figure*}
\centerline{\includegraphics[width=17.5cm]{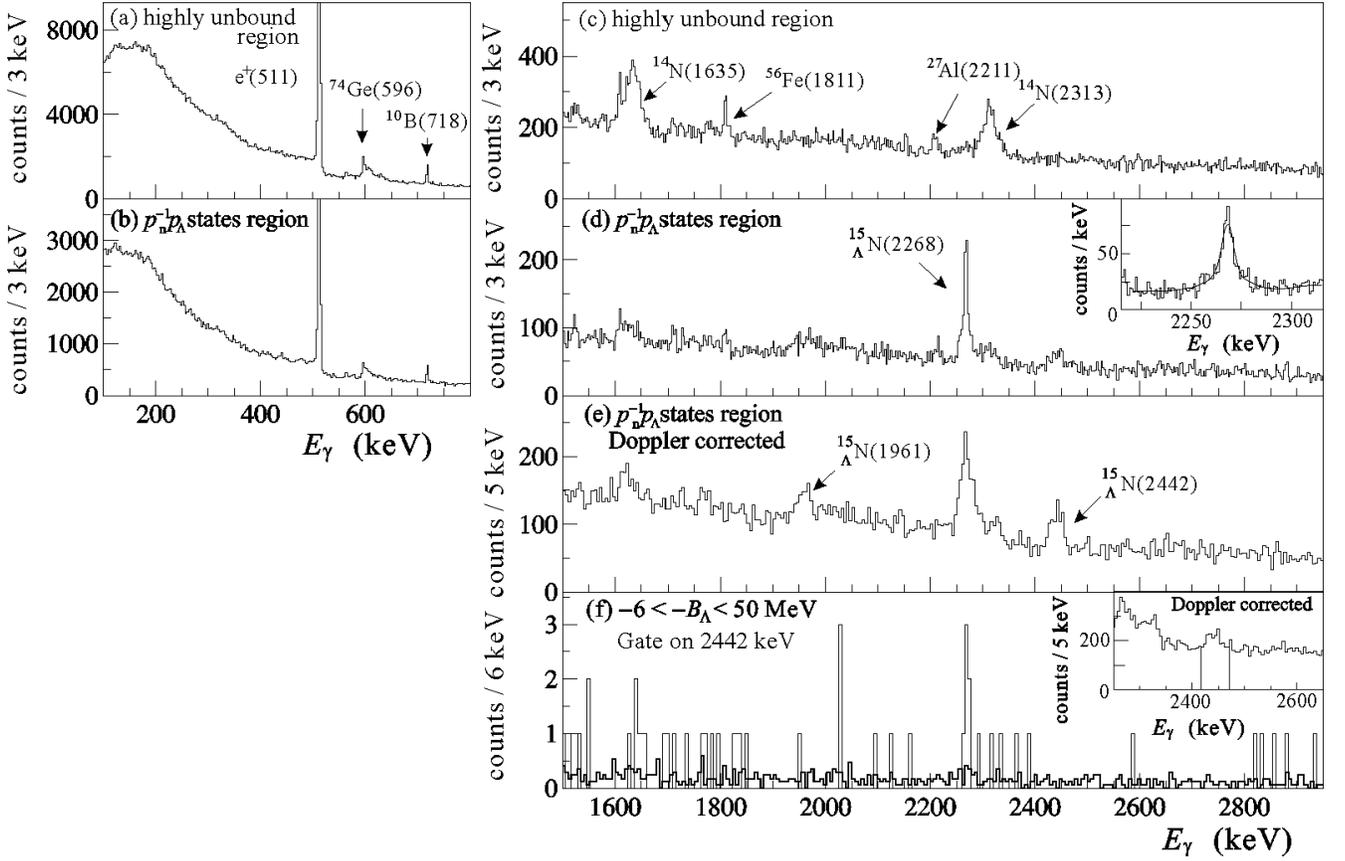}}
\caption{Mass-gated \ggg -ray spectra from the \Kpi\ reaction
  on the 20 g/cm$^2$ H$_2$O target. (a) and (c) are plotted for the
highly-unbound region ($-B_\Lambda > 50$ MeV), and (b), (d) and (e) are
	 for the \pp -states region. (e) is the spectrum with
event-by-event Doppler-shift corrections ($E_\gamma^C$) applied to (d).
     Three \ggg -ray peaks, a narrow peak at 2268 keV in (d) and
	 Doppler-broadened peaks at 1961 and 2442 keV in (d),
		 which become narrower peaks in (e),
	    are assigned as \ggg \ rays from \lam{15}{N}.
	      (f) is a \ggg -\ggg \ coincidence spectrum
	      plotted  for $-6 < -B_\Lambda < 50 $ MeV.
 The solid line is the spectrum of \ggg \ rays accompanied by another
\ggg \ ray  with $E_\gamma^C =2442 \pm 25$ keV, as shown in the inset.
The shaded region in (f), which shows the background level, is the spectrum of 
\ggg \ rays accompanied by another \ggg \ ray with 
$E_\gamma^C = 2600$ -- 4000 keV, and with the counts scaled to 
the \ggg -ray counts at 2442 $\pm$ 25 keV.}
\label{nigam}
\end{figure*}

 Figure \ref{nigam} shows the mass-gated \ggg -ray spectra measured in 
coincidence with the \EL{16}{O}\Kpi\ reaction; (a) and (c) are  plotted 
for highly unbound region ($ -B_\Lambda > 50$ MeV), (b), (d) and (e) are 
for the \pp  -states region ($-12 < -B_\Lambda < 14$ MeV).
Event-by-event Doppler-shift corrections are applied for (e). 
The highly-unbound and \pp -states regions are defined in Fig.~\ref{oxblam}.

 A prominent peak is found at 2268 keV in spectrum (d) which is gated on 
the \pp -states region. The \ggg -ray peak is taken to be hypernuclear
since there are no \ggg -ray transitions of this energy in ordinary 
nuclei with $A \le$ 16. The \ggg -ray peak has two components, namely a 
narrow part and a Doppler-broadened part, which indicates that the 
lifetime of this transition is close to the stopping time of $\sim$ 1 ps
for the hypernucleus in the target. There are four candidates for 
secondary hypernuclei with thresholds below the \lam{16}{O}($0^+_2$) state,
namely \lam{15}{N}, \lam{15}{O}, \lam{13}{C}, and \lam{12}{C} 
(see Fig.~\ref{onrelate}). 
The long lifetime of the 2268-keV \ggg \ ray suggests that the corresponding 
core transition is also slow leaving the $0^+;1$, first-excited state of 
\EL{14}{N} ($E_x= 2.31$ MeV, $\tau =98$ fs) as the only candidate. 
Moreover, the excitation energies of members of the first-excited doublets in
in \lam{12}{C} and \lam{13}{C} are known to be 2.51 MeV for
\lam{12}{C}($1^-_2$)~\cite{hotchi01} (2.63 MeV from KEK 
E336~\cite{hashtam06}) and 4.88 MeV for \lam{13}{C}($3/2^+$)~\cite{kohri02} 
(4.85 MeV~\cite{hashtam06}).
The unknown first-excited state of \lam{15}{O} must
be based on the 5.17-MeV $1^-$ state of \EL{14}{O} and the $0^+_2$ state 
of \lam{16}{O} can't decay to a state at this high an energy.
Therefore, the 2268-keV \ggg  -ray peak is attributed to \lam{15}{N}.
In addition, two broad peaks at 1960 and 2440 keV can be seen in (d) and
they become square-shaped narrower peaks in (e) after the Doppler-shift 
correction. These peak shapes are consistent with the result of the 
simulation for \lam{15}{N} described in Sect.~\ref{shift-correction} 
[see Fig.~\ref{ngamsim} (a) and (b)] while, according to the simulation,
\ggg -ray peaks from \lam{12}{C} and \lam{13}{C} can't become narrower.
Therefore, these two \ggg  \ rays are also attributed to transitions 
in \lam{15}{N}. On the other hand, a narrow \ggg -ray peak observed at 
2215 keV can be attributed to a part of the broad \ggg -ray peak from 
\EL{27}{Al}($n,n'$\ggg) because it is clearly enhanced by the 
``delayed'' TDC cut [see Fig.~\ref{tdccut}].   

As shown in Fig.~\ref{nigam} (b), no prominent peak corresponding 
to the ground-state spin-flip M1 transition is observed in the region 
from 100 keV to 700 keV. The sensitivity for such a peak 
and the ground-state doublet spacing will be discussed later 
(Sect.~\ref{sec:gsdoublet15}).

We made  fits for the Doppler-shift corrected \ggg -ray spectrum
using the simulated peak shape from Fig.~\ref{ngamsim}(a).
We obtained energies of $1960.7 ^{+ 1.2}_{-1.7} \pm 1.7 $ keV and 
2442.0 $^{+0.7}_{-1.7}$ $\pm$1.7 keV, with corresponding yields of 
190 $^{+30}_{-36} $ $\pm$ 5 and 313 $\pm$ 35 $\pm$ 5 counts.
For the narrow peak, we obtained the energy of 2267.6 $\pm$ 0.3 $\pm$ 
1.5 keV and a yield of 744 $\pm$ 39 $\pm$ 15 counts.
The detailed fitting procedure for the 2268-keV \ggg \ ray is described 
later. The measured  \ggg -ray energies correspond to transition 
energies of 1960.8, 2267.8 and 2442.3 keV. Taking into account the 
energy dependence of the \ggg -ray efficiency, the  relative \ggg -ray 
intensities for the \pp -states region were found to be
\begin{eqnarray} 
\label{Nigamma}
I_\gamma(1961): I_\gamma(2268):I_\gamma(2442) \nonumber \\
=0.23 ^{+0.04}_{-0.05} : 1 : 0.45 \pm 0.05 .  
\end{eqnarray}
  
 Figure \ref{nigam} (f) shows the spectrum of \ggg \ rays emitted in 
coincidence with another \ggg \ ray with a Doppler-corrected energy
of $E_\gamma^C=2442 \pm 25$ keV  for the \lam{16}{O} mass 
range of $-6 < -B_\Lambda < 50$ MeV. The shaded region in (f) shows 
the expected background, which is derived from the 
spectrum of \ggg \ rays accompanied by another \ggg \ ray with
$E_\gamma^C = 2600 - 4000$ keV for the same mass region,
with the counts scaled to the counts in the 2442 $\pm$ 25 keV  gate
corresponding to the shaded region in the inset (f).
The mass region $-6 < -B_\Lambda < 50$ MeV
was set to select  the $0^+_2$ state ($-B_\Lambda = 4$ MeV) and higher states. 
A peak with 6 counts appeared at 2268 keV, the energy of one of the other 
\ggg \ rays from \lam{15}{N}. The probability for 6 events to appear in 
the narrow region specified by 2268 $\pm$ 12 keV from a random fluctuation 
of the background of 1.2 $\pm$ 0.3 counts was estimated to be 0.4 \% using
 a Poisson distribution. Thus the possibility of a background 
fluctuation was rejected. Since the peak shapes indicate that the lifetime 
of the 2442-keV transition is shorter than that the 2268-keV transition,
the decay chain \ggg (2442 keV) $\to$ \ggg (2268 keV) was determined.
The yield of the 2442-keV \ggg \ ray for the $-6 < -B_\Lambda < 50$ MeV 
region was obtained to be $330 \pm 47 \pm 10$ counts and the expected  
yield of the 2268-keV \ggg \ ray in coincidence is estimated to be 
4.6 counts if no other transition strongly competes  with the 2268-keV 
transition. Therefore, the 6 counts with 1.2 $\pm$ 0.3 counts background 
is consistent with the expected coincidence yield. On the other hand, 
no excess of events was found in the \ggg-\ggg \ coincidence 
spectrum with the 1961-keV \ggg \ ray, consistent with the lower yield
of the 1961-keV \ggg \ ray.

\subsubsection{Angular distributions}
\label{sec:angdist15}

\begin{figure*}[hbt]
\begin{minipage}[h]{12cm}
\centerline{\includegraphics[width=11.71cm]{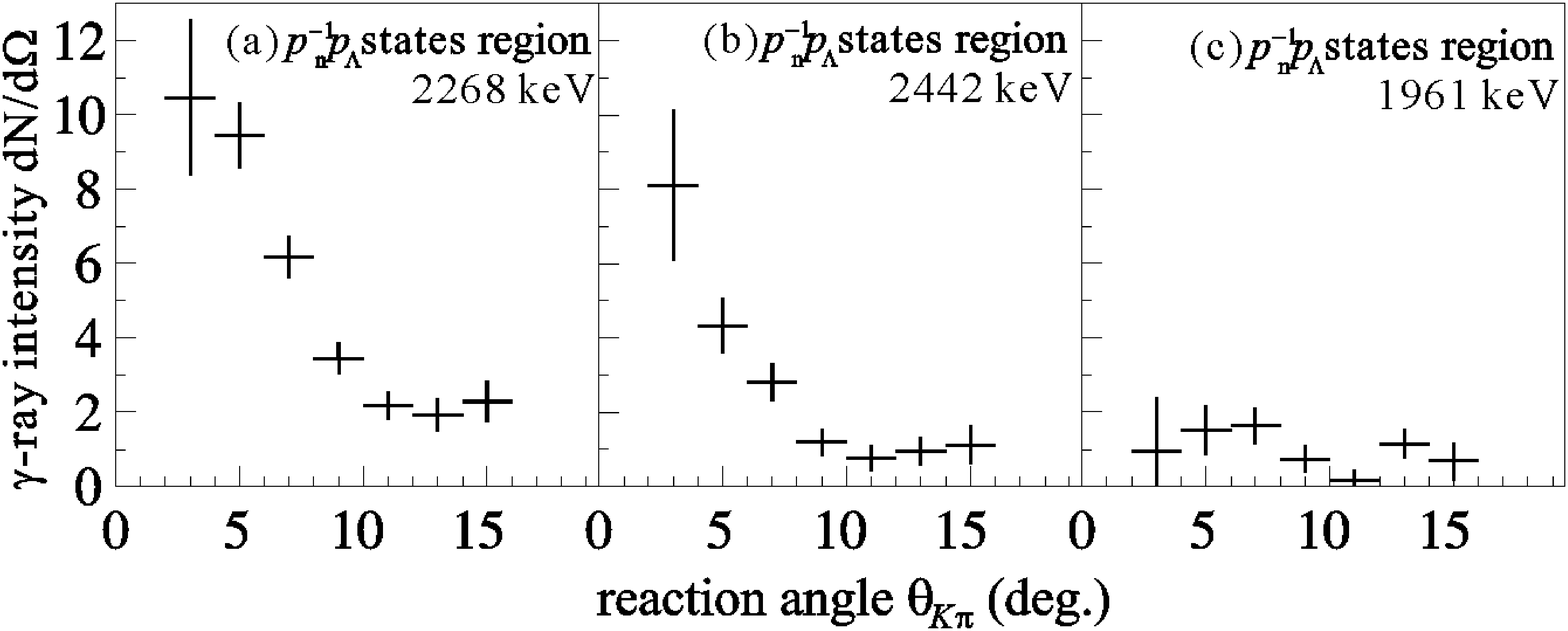}}
\end{minipage}
\begin{minipage}[r]{5.5cm}
\caption{Reaction angle ($\theta_{K\pi}$) dependence of the  
\lam{15}{N} \ggg \ ray intensities measured in the \EL{16}{O}\Kpi\ reaction
for the \pp -states mass region ($-12 < B_\Lambda < 14$ MeV) in \lam{16}{O}.
The panels show the relative intensities of 
(a) 2268-keV, (b) 2442-keV, and (c) 1961-keV \ggg \ rays.
Relative intensities were obtained by dividing the measured \ggg -ray 
counts by the spectrometer acceptance (Fig.~\ref{acceptance}). The relative 
\ggg -ray efficiencies are normalized to that of the 6550-keV \ggg \ ray.}
\label{angdist}
\end{minipage}
\end{figure*}

 Figure \ref{angdist} (a--c) shows the reaction angle ($\theta_{K\pi}$) 
dependence of the  \lam{15}{N} \ggg -ray intensities for the \pp -states 
region  in the \EL{16}{O}\Kpi\ reaction. The \ggg -ray intensities were 
obtained from the measured \ggg -ray counts divided by the spectrometer 
acceptance (Fig.~\ref{acceptance}) and normalized to the relative 
efficiency for the 6550-keV \ggg\ ray. The panels of the figure show the  
distributions for the (a) 2268-keV, (b) 2442-keV, and (c) 1961-keV \ggg \ 
rays. The spectra (a) and (b) show forward peaking angular distributions
with a significant intensity remaining at large angles 
($\theta_{K\pi} > 10^\circ $). This indicates that the \lam{15}{N} 
excited states de-excited by the 2268- and 2442-keV \ggg \ rays are 
produced mainly by proton emission from \lam{16}{O} states excited by
$\Delta L=0$ transitions with some contribution from states produced
by $\Delta L=2$ transitions. On the other hand, the \lam{15}{N} excited 
state leading to the 1961-keV \ggg \ deexcitation is mainly produced 
from non-substitutional states excited  by $\Delta L \ge 1$
transitions. 

\subsubsection{Level assignments}
\label{sec:assign15}

 The partly broadened shape of the 2268-keV \ggg -ray peak
indicates a lifetime comparable to the stopping time of $\sim 1$ ps.
Since the $^{14}$N($0^+;1$ $\to$ $1^+;0$) transition has 
a lifetime of 0.10 ps and an energy of 2313 keV,
the 2268-keV \ggg  \ ray is attributed to  
a \lam{15}{N} transition corresponding to this core transition,
namely, from the $1/2^+;1$ state to one of 
the ground-state doublet members $(3/2^+, 1/2^+)$.
In the weak-coupling limit, the $B(M1)$ values for the decay to the 
$3/2^+$ and $1/2^+$ states would be in the ratio of $2\!:\!1$. 
However, taking into account the level mixing,
they are calculated to be in the ratio of $10\!:\!1$ \cite{millener05}. 
Therefore,  the 2268-keV \ggg \ ray is taken to correspond to 
the $1/2^+;1\to3/2^+_1$ transition.

The lowest state in \EL{14}{N} which decays to the $0^+;1$ state
(the core level of the $1/2^+;1$ state in \lam{15}{N})
is the 3948-keV $1^+$ state and it gives rise to the $3/2^+_2$, 
$1/2^+_2$ doublet. Therefore, the initial state for the  2442-keV 
\ggg -ray transition is most likely to be one of these doublet members. 
The $\Delta L=0$ dominance shown in the angular distribution
(Fig.~\ref{angdist} (b)) indicates that the 2442-keV \ggg \ ray
mostly stems from the \lam{16}{O}($0^+_2$) state or the $0s$-hole state
(see Fig.~\ref{onrelate}). The TISM calculation \cite{majling92} suggests that 
the \lam{15}{N}($1/2^+_2$) state, rather than the \lam{15}{N}($3/2^+_2$) 
state, is dominantly produced from the \lam{16}{O}($0^+_2$) state.
On this basis, the 2442-keV \ggg \ ray is likely to be the $1/2^+_2 
\to 1/2^+;1$ transition.
  
All the excited states higher than the $0^+;1$ state in \EL{14}{N} 
decay mainly to the $0^+;1$ state or the $1^+;0$ ground state.
Since  no other \ggg \ rays from \lam{15}{N} are observed,
the 1961-keV \ggg  -ray transition is also likely to be 
a transition decaying to the $1/2^+;1$ state. The $\Delta L \ne 0$ 
dominance in the angular distribution   (Fig.~\ref{angdist} (c)) 
implies that the 1961-keV \ggg \ ray is predominantly emitted from 
a state populated by proton emission from the $2^+$  states of 
\lam{16}{O}($2^+$). The TISM calculation suggests that the $2^+$ states 
in \lam{16}{O} decay to the $3/2^+_2$ state rather than the $1/2^+_2$ 
state. On this basis, the 1961-keV \ggg \ ray is likely to be   
the $3/2^+_2 \to 1/2^+;1$ transition.

 As described above, the $\theta_{K\pi}$ distributions of the 
2442-keV and 1961-keV \ggg \ rays imply that the $1/2^+$ state
is the upper member and the $3/2^+$ state is the lower member of
the excited doublet. Because the 3948-keV $1^+;0$ state of $^{14}$N
is mainly $^3S$ in nature (see Sect.~\ref{sec:discussion}),
the upper-doublet spacing is dominantly due to the spin-spin 
interaction term ($\Delta$) and for $\Delta >0$ the  
the spin anti-parallel state becomes the lower level, 
as for \lamb{7}{Li}($1/2^+$). The contradiction between the 
implications of the $\gamma$-ray yields as a function of pion 
scattering angle and the underling structure of \lam{15}{N} will 
be discussed in Sect.~\ref{sec:discussion}. 

\subsubsection{Missing-mass spectra}
\label{sec:missingmass15}

\begin{figure}[h]
\centerline{\includegraphics[width=8.5cm]{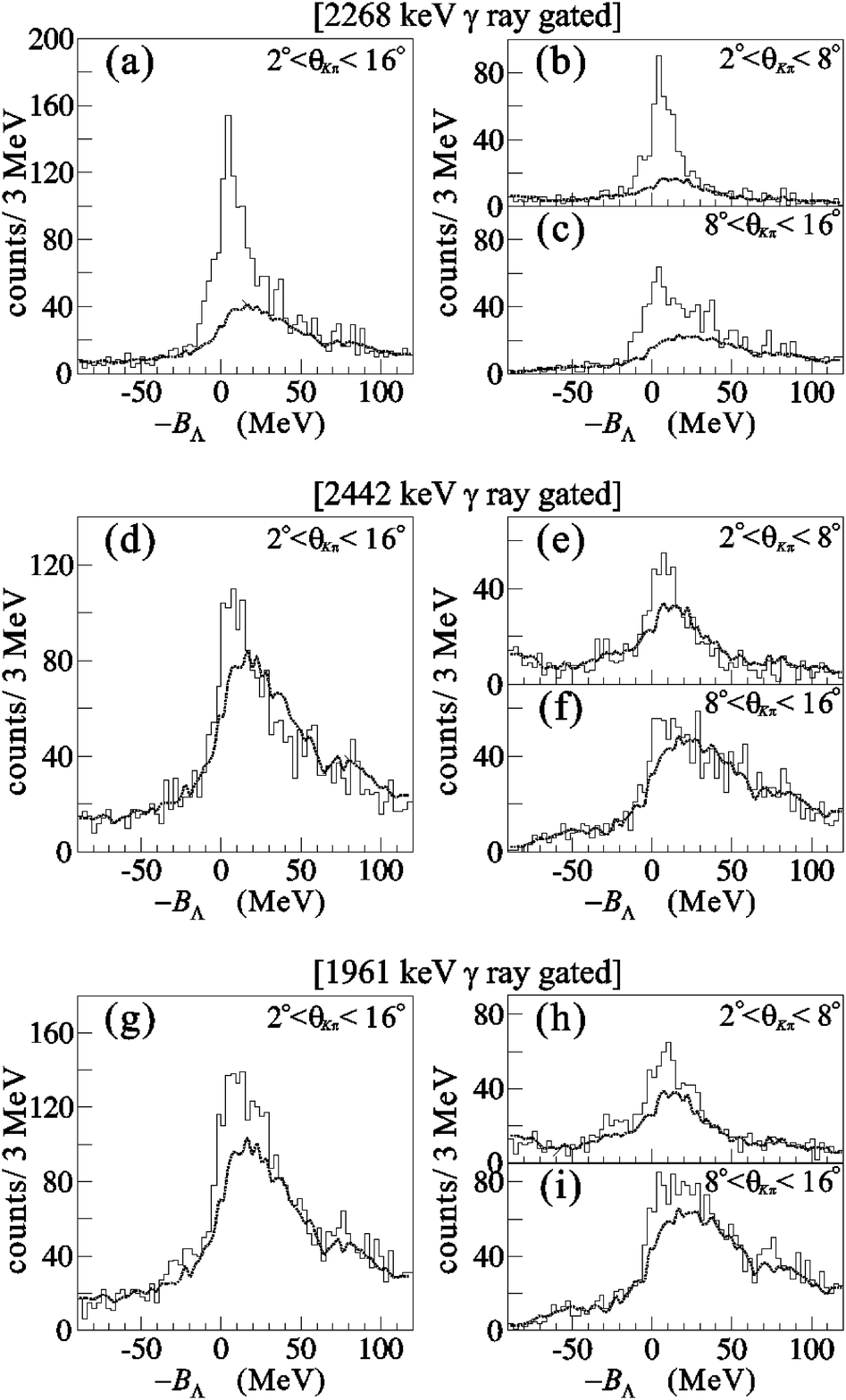}}
\caption{Missing-mass spectra for the \EL{16}{O}\Kpi\ reaction 
as a function of $\Lambda$ binding energy ($B_\Lambda$).
The solid lines are plotted for those events accompanying \ggg  \ rays
in the energy range $E_\gamma ^M=2268 \pm $ 8 keV for (a), (b), and (c),
$E_\gamma ^C =2442 \pm 16$ keV for (d), (e), and (f),
and $E_\gamma ^C =1961 \pm 16$ keV for (g), (f) and  (i).
The reaction angles covered are
$2^\circ < \theta_{K\pi} < 16^\circ$ for (a), (d), and (g),
$2^\circ < \theta_{K\pi} < 8^\circ$  for (b), (e), and (f),
$8^\circ < \theta_{K\pi} < 16^\circ$ for (c), (f), and (i).
The dashed lines show the expected background shapes for the 
corresponding reaction angles, determined from the events accompanying 
\ggg  \ rays with  energies from 2500 keV to 3000 keV after scaling 
to fit the background region ($-90< -B_\Lambda < -30$ MeV).}
\label{blam2268}
\end{figure}

 Figure \ref{blam2268} shows missing-mass spectra, plotted as a 
function of $\Lambda$ binding energy ($B_\Lambda$), for the three 
$\gamma$ rays attributed to \lam{15}{N} in the \EL{16}{O}\Kpig\ 
reaction. The energy and angular ranges are specified in the caption.

By fitting the excess over the background with a Gaussian, 
the peak widths gated on the 2268-keV \ggg \ ray were found
to be $21\pm 2$, $19\pm 2$, and $23\pm 3$ MeV (FWHM) from 
Figs.~\ref{blam2268} (a), (b), and (c), respectively. These widths 
are significantly wider than the mass resolution. The natural widths 
of the \pp \ states and the separations between the $0+$ and $2^+$ 
states in each group of levels (see Fig.~\ref{onrelate})
are much smaller than the mass resolution (see Fig.~\ref{onrelate})
\cite{bruckner78,hashimoto98}, suggesting that the $1/2^+;1$ state in 
\lam{15}{N} is not produced from a single group of states in \lam{16}{O}.
By comparing (b) with (c), more events are clearly seen at around 
$20 - 40$ MeV in (c) showing  that the $1/2^+;1$ state of \lam{15}{N}
is fed by proton emission decay from several states of \lam{16}{O}
produced by $\Delta L > 0$ transitions.

On the other hand, the widths from Figs.~\ref{blam2268} (d), (e), and (f) 
were  17 $\pm$ 2, 17 $\pm$ 3 and 17 $\pm$ 5 MeV (FWHM), respectively.
This suggests that the state which emits 2442 keV \ggg \ ray is dominantly 
produced from the $0^+_2$ state  and its partners in \lam{16}{O}.

The widths from Figs.~\ref{blam2268} (g), (f), and (i) were  
26 $\pm$ 8, 26 $\pm$ 8 and 18 $\pm$ 9 MeV (FWHM), respectively,
and no information on the population from states of \lam{16}{O} 
can be obtained.

\subsubsection{Lifetime analysis}
\label{sec:lifetime15}

The lifetime of the $1/2^+;1$ state was derived from the shape of the 
peak by fitting with the simulated shape. Since the recoil velocity is 
affected not by the absolute masses of \lam{16}{O} and \lam{15}{N} but 
by the Q value of the proton decay, it is necessary to know the masses 
of the \lam{16}{O} and \lam{15}{N} states contributing to the decay.
In the simulation, we used the excitation energies of the $0^+, 2^+$ states 
and the proton emission threshold as shown in Fig.~\ref{onrelate}.
We assumed that all the excited states in \lam{15}{N} are produced directly 
from the excited $0^+$ and $2^+$ states at 17 MeV and that the
1961- and 2442-keV \ggg \ rays  are from decays to the $1/2^+;1$ state 
with the intensities in Eq.~(\ref{Nigamma}). Here, the lifetimes of the 
upper doublet states were assumed to be much shorter than that of the 
$1/2^+;1$ state and their effect was neglected. This assumption is 
justified because the core state of the doublet [\EL{14}{N}($1^+_2$)] has 
a short lifetime (7 fs) compared to the scale set by the stopping time.
The measured $\theta_{K\pi}$ distribution from  Fig.~\ref{angdist}~(a) 
was used (cf. Fig.~\ref{dwia}). In addition, the reaction angle was 
selected in the region from $2^\circ$ to $8^\circ$ to reduce the 
contributions from the non-substitutional states evident in the $20 - 40$
MeV region of Fig.~\ref{blam2268}~(c). The uncertainty in the 
excitation energies of states in \lam{16}{O} from previous experiments 
was found to have a  negligible effect. The systematic error on the 
lifetime stems mainly from lack of knowlwdge of the decay ratio  
of the $0^+_1$ and $0^+_2$ states to the $1/2^+;1$ state ($\pm 0.2$ ps).
Uncertainty in the response function for the \ggg -ray peak at this 
energy was found to also contribute to the error in the lifetime 
($\pm 0.1$ ps). Then, the lifetime obtained for the $1/2^+;1$ state
is $1.5\pm 0.3$ (stat.) $\pm 0.3$ (syst.)  ps. This lifetime 
is 15 times longer than that of the $0^+;1$ core state and longer
than the theoretically predicted value of 0.5 ps \cite{millener05}.
The reason for this dramatic increase in the lifetime of the 
hypernuclear state is discussed in Sec.~\ref{sec:discussion}.

 We also examined the lifetimes of the initial states of the 
1961- and 2442-keV \ggg \ rays by  fitting the spectrum without 
Doppler-shift correction. In the simulation of the peak shape,
we assumed that the $\theta_{K\pi}$ distributions were the same as 
that of the 2268-keV \ggg \ ray. 
Then the upper limits (68\% C.L.) on the lifetimes were found to be  
$\tau < 0.2$ ps for the 1961-keV \ggg \ ray and
$\tau < 0.3$ ps  for the 2442-keV \ggg \ ray.

\subsubsection{Ground-state doublet spacing}
\label{sec:gsdoublet15}

 In Sec.~\ref{sec:assign15}, the 2268-keV $\gamma$ ray was taken to
correspond to the $1/2^+;1\to 3/2^+_1$ transition based on
the theoretical expectation that this transition should be by far the
stronger of the two possible $\gamma$ rays de-exciting the $1/2^+;1$ 
level. The energy of the $1/2^+;1\to 1/2^+_1$ transition would 
determine the ground-state doublet spacing but no candidate is 
observed. In addition, no low-energy $\gamma$-ray peak corresponding to 
to the ground-state doublet, spin-flip transition 
($1/2^+_1 \to 3/2^+_1$ or $3/2^+_1 \to 1/2^+_1$) is observed.
In these circumstances, nothing can be said about the ground-state
doublet spacing if the $1/2^+$ state is the upper member of the 
doublet as the shell-model calculations predict (see 
Sec.~\ref{sec:discussion}).

 If, on the other hand, the $3/2^+_1$ state is the upper level of the 
doublet, as expected in the the $jj$-coupling limit~\cite{millener85} given 
that \lam{16}{O} has a $0^-$ ground state, an upper limit can be put on
the doublet spacing. The B(M1) for the $3/2^+_1 \to 1/2^+_1$ transition
can be estimated quite accurately in the weak-coupling limit~\cite{dalitz78}
as
\begin{equation}
 B(M1;3/2^+_1 \to 1/2^+_1) = (g_\Lambda - g_c)^2/4\pi\ \mu_N^2,
\end{equation}
where $g_c = 0.4038$~\cite{ajz91} and $g_\Lambda = -1.226$~\cite{pdg}.
Then, $\tau(3/2^+) = 0.269\,E_\gamma^{-3}$ ps with $E_\gamma$ in MeV. 
Taking into account the \ggg -ray efficiency and the competition between 
electromagnetic and weak decay ($\tau_\textrm{weak}\sim 200$ ps for 
most nuclei~\cite{kameoka05,grace85,bhang98}) for the $3/2^+_1$ state, 
the sensitivity of the experiment is high enough to observe 
a \ggg -ray peak with $E_\gamma > 100$ keV even if the $3/2^+_1$  
state was not produced from \lam{16}{O} directly. Then we can conclude 
that the ground-state doublet spacing in \lam{15}{N} is
\begin{equation} 
E(3/2^+_1)-E(1/2^+_1) < 100  ~\textrm{keV}.
\end{equation}

\subsubsection{The $1/2^+;1 \to 1/2^+_1$ transition}
\label{sec:branch15}

\begin{figure}[th]
\centerline{\includegraphics[width=8.5cm]{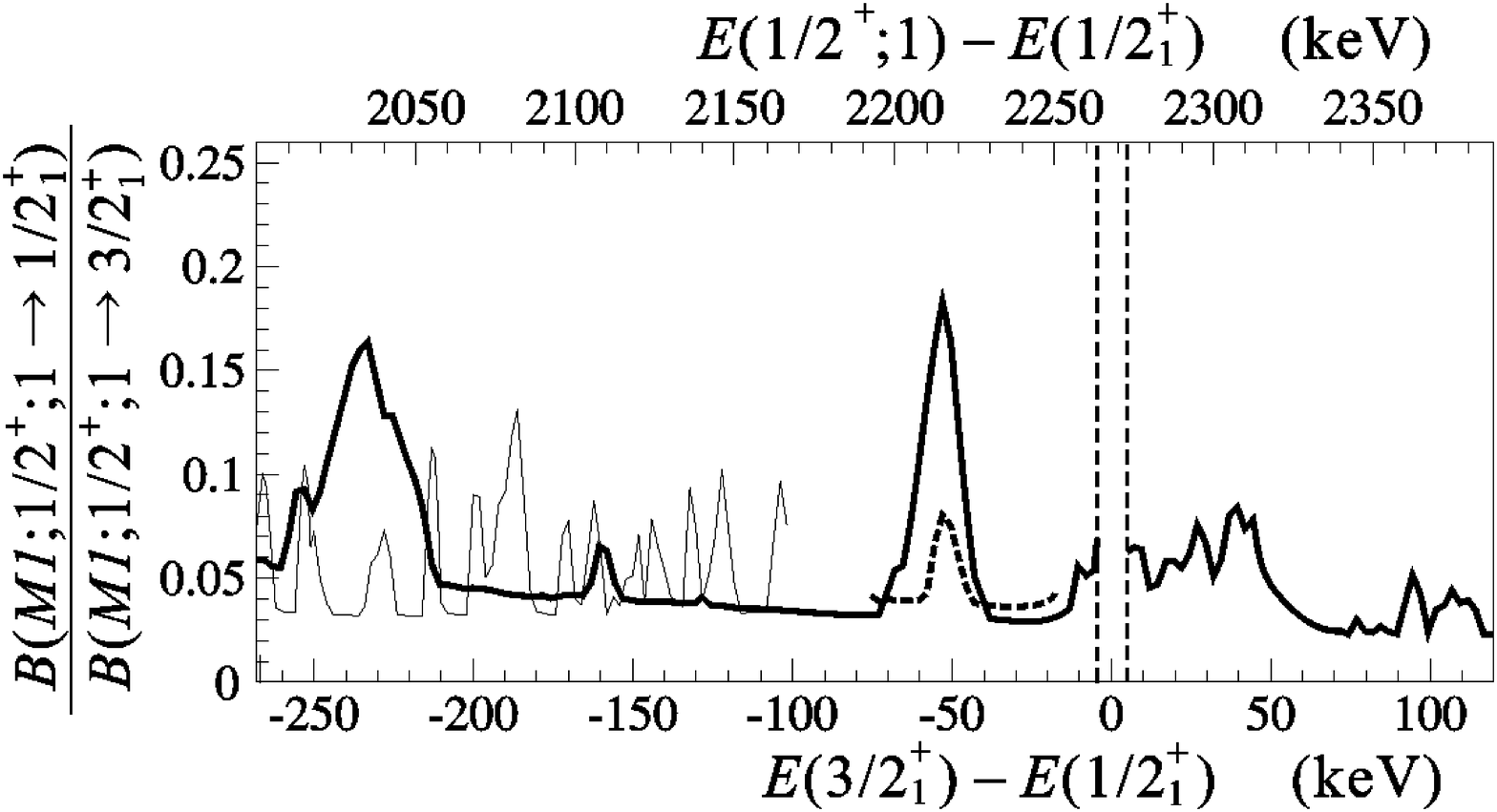}}
\caption{Upper limit for the $B(M1)$ ratio between the 
$1/2^+;1 \to 1/2^+_1$ and $1/2^+;1 \to 3/2^+_1$ transitions as a 
function of  the ground-state doublet spacing, $E(3/2^+_1)-E(1/2^+_1)$, 
and the $1/2^+;1 \to 1/2^+_1$ transition energy. The thick line is the 
result of a peak search for the $1/2^+;1 \to 1/2^+_1$ transition obtained 
by fitting the spectrum from 2000 keV to 2370 keV [Fig.~\ref{nigam} (d)]. 
The dashed line is from the same procedure but with the ``prompt'' TDC cut. 
The thin line is the result of a peak search for the $1/2^+_1 \to 3/2^+_1$ 
transition obtained by fitting the spectrum from 100 keV to 270 keV
[Fig.~\ref{nigam} (b)].}
\label{bm1r}
\end{figure}

 As noted previously, no prominent \ggg -ray peak corresponding to 
the $1/2^+;1 \to 1/2^+_1$ transition has been observed.
An upper limit for the $B(M1)$ ratio between the $1/2^+;1 \to 1/2^+_1$ 
and the $1/2^+;1 \to 3/2^+_1$ transitions as a function of the 
ground-state doublet spacing can be obtained from the \ggg -ray spectrum. 
This is shown in Fig.~\ref{bm1r} where the shaded region shows the 
allowed range of the $B(M1)$ ratio. The ratios in Fig.~\ref{bm1r} 
were obtained from the upper limit of the \ggg -ray yield taking into 
account the relative \ggg -ray efficiencies.

The thick line shows the upper limit of the $B(M1)$ ratio
obtained by fitting the spectrum from 2000 to 2370 keV
assuming that the $1/2^+;1 \to 1/2^+_1$ transition has 
the same \ggg -ray peak shape as the $1/2^+;1 \to 3/2^+_1$ transition.
The dashed line is obtained by fitting the spectrum with 
the ``prompt'' TDC cut to reduce the \EL{27}{Al}$(n,n')$ \ggg \ ray at 
2211 keV. The thin line is obtained by fitting the spectrum from 100 keV 
to 270 keV, assuming that the $1/2^+_1 \to 3/2^+_1$ \ggg \ ray has the 
original response function because the lifetime of the $1/2^+_1$ state  
(half the estimate for the $3/2^+_1$ level in the previous subsection)
is much longer than the stopping time and with the estimates for the
electromagnetic and weak-decay lifetimes taken into account.

 The $B(M1)$ ratio between the $1/2^+;1 \to 1/2^+_1$ and 
$1/2^+;1 \to 3/2^+_1$ transitions is calculated to be $10\!:\!1$ 
with the level mixing taken into account \cite{millener05} using
the $\Lambda N$ parameters from the first analysis of the \lam{16}{O}
$\gamma$-ray data~\cite{ukai04}. As shown in Fig.~\ref{bm1r}, either
the doublet spacing is less than 5 keV or the upper limit on the B(M1)
ratio is $\sim 0.09$ (the calculation puts the $1/2^+_1$ level
$\sim 100$ keV above the $3/2^+_1$ level). The interpretation of
this result is discussed in detail in Sec.~\ref{sec:discussion}.
Basically, small $1^+_2\times s_\Lambda$ admixtures in the wave 
functions of the ground-state doublet members introduce strong 
destructive interferences from the strong $1^+_2;0\to 0^+;1$ M1 core 
transition into the $\gamma$ decays from the $1/2^+;1$ level of 
\lam{15}{N}.

\subsection{Summary of results}
\label{sec:results-summary}
 
\begin{figure}[h]
\centerline{\includegraphics[width=8.0cm]{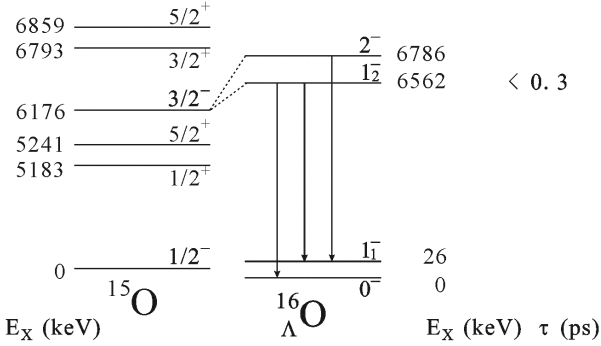}}
\caption{Experimentally determined level scheme of \lam{16}{O}
and observed \ggg -ray transitions.
The corresponding level 
scheme of \EL{15}{O} is also shown.
}
\label{oxlevel}
\end{figure}

Figure \ref{oxlevel} shows the experimentally determined level 
scheme of \lam{16}{O}. As shown in the figure, the excitation 
energies and spin ordering of both ground and 6.7-MeV excited 
doublets have been determined. This is the first determination 
of the spin-ordering and the spacing of a $p_{1/2}$-shell 
hypernuclear ground-state doublet. In addition, this is the first 
observation of a spin-flip state (the $2^-$ state) 
that is directly produced via the \Kpi\ reaction. 

\begin{figure}[h]
\centerline{\includegraphics[width=8.0cm]{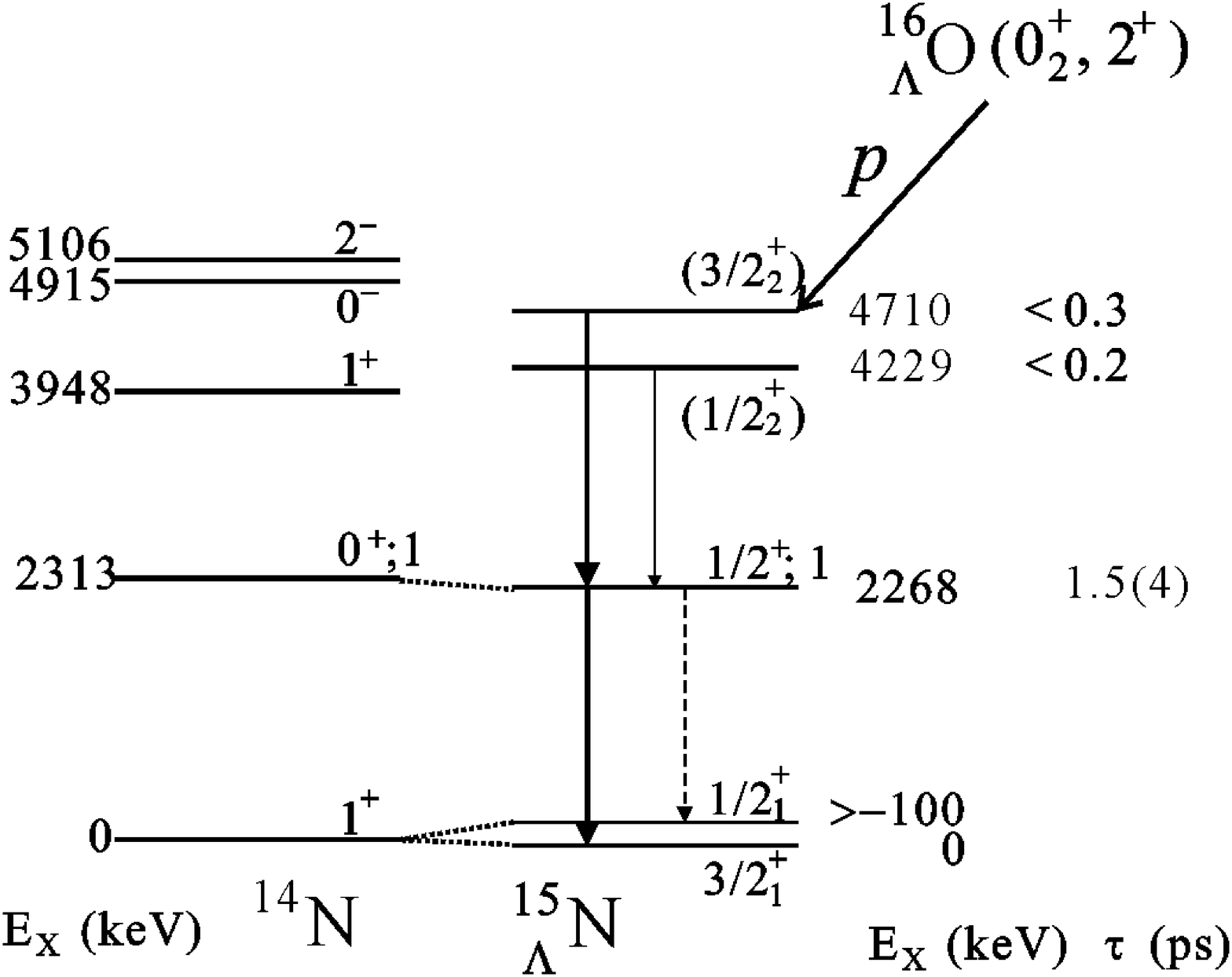}}
\caption{Experimentally determined level scheme of \lam{15}{N}
and observed \ggg -ray transitions (solid arrows). 
The corresponding level scheme 
of the \EL{14}{N} core nucleus is also shown. Since the spin ordering 
of the ground-state doublet is not determined,
the excitation energies of \lam{15}{N} are given as level spacings
from the $3/2^+_1$ state.
The existence of the 4229-keV excited state was not 
experimentally determined but the 1961-keV \ggg \ ray is 
likely a transition  to the $1/2^+;1$ state.
}
\label{nilevel}
\end{figure}

Figure \ref{nilevel} shows the level scheme of \lam{15}{N} 
together with the corresponding core levels of $^{14}$N~\cite{ajz91}.
Since the spin ordering and spacing of the ground-state doublet 
were not experimentally determined, the excitation energies are 
given as level spacings from the $3/2^+_1$ state.
The existence of the 4710-keV excited state was determined from 
the coincidence of the 2442-keV \ggg \ ray with the 2268-keV 
\ggg \ ray. On the other hand, while the  existence of the 4229-keV 
excited state was not unambiguously determined, the 1961-keV 
\ggg \ ray is likely a transition  to the $1/2^+;1$ state. 

The 
observed \ggg -ray transitions are summarized in Table~\ref{ngam}.

\begin{table*}
\caption{\label{ngam}
Observed hypernuclear \ggg \ rays by the \Kpi reaction.
Intensities are normalized to the largest yield \ggg \ rays.
The 2442-keV and 1961-keV \ggg \ ray transitions in
\lam{15}{N} are not experimentally determined but most likely assignments 
are shown. }
\begin{ruledtabular}
\begin{tabular}{cccccc}
Gated mass [MeV] & $E_\gamma$ (keV) & Counts& Relative intensity 
& \lam{A}{Z} & ($J_i;T_i\to J_f;T_f$) \\
\hline 
$-17<-B_\Lambda < 3$ &$6560.3\pm1.1\pm1.7$ &$183\pm16\pm5$ & 100 &
 \lam{16}{O}&$1^-_2\to0^-$ \\
 & $6533.9\pm1.2\pm1.7$ &$127\pm16\pm5$ & $69\pm11\pm10$ &
\lam{16}{O}&$1^-_2\to1^-_1$\\
 & & & & & \\
$-13<-B_\Lambda < -1$ &$6758 \pm 4 \pm 4$   & $21.0^{+7.2}_{-6.5}$ & 
 & \lam{16}{O}&$2^-\to 1^-_1$\\ 
 & & & & & \\
 &$2442.0^{+0.7}_{-1.7} \pm 1.7$  & $313\pm35 \pm 5$ & $49\pm 5$
 & \lam{15}{N} & 
~~~~~~~~~~~$\to 1/2^+;1$ \\
 $-12 <-B_\Lambda < 14$ 
 &$1960.7^{+1.2}_{-1.7} \pm 1.7$  &  $190 ^{+30}_{-36} \pm 5$&  $23\pm5$ 
& \lam{15}{N} & 
~~~~~~~~~~~($\to 1/2^+;1$) \\
 &$2267.6\pm 0.3 \pm 1.5$  &  $744\pm 39 \pm 15$ & 100  
 & \lam{15}{N} & $1/2^+;1\to3/2^+;0$ \\
\end{tabular}
\end{ruledtabular}
\end{table*}

\section{Discussion}
\label{sec:discussion}

There are some uncertainties concerning the level assignments for
the upper doublet members in both hypernuclei. In \lam{16}{O}, the
6758-keV transition assigned as originating from the $2^-$ level has a 
statistical significance of $3\sigma$ (Sec.~\ref{assign16}). Also,
it cannot be completely ruled out that it comes from a positive-parity
state (see Fig.~\ref{oxlevel}) populated by a weak non-spin-flip,
$\Delta L =2$ transition in the \Kpi\ reaction. In \lam{15}{N},
the tentative assignments for the upper doublet shown in 
Fig.~\ref{nilevel} are based on shell-model calculations using 
positive $\Delta$ values like that in Eq.~(\ref{eq:param7}) (and in 
analogy to the ground-state doublet in \lamb{7}{Li}). The $\gamma$-ray 
yields as a function of the reaction angle $\theta_{K\pi}$  
(Sec.~\ref{sec:assign15}) suggest an inverted order for this doublet.
However, the interpretation of these yields depends on an
approximate $1\hbar\omega$ shell-model calculation \cite{majling92}
to estimate the population of states in \lam{15}{N} via proton emission 
from the $0^+$ and $2^+$ states of \lam{16}{O} (see the end of 
Sec.~\ref{sec:intro-exp}). These predictions
depend on small amplitudes [the $\beta$ in Eq.~(\ref{config})] 
in the \lam{16}{O} wave functions that are difficult to estimate reliably.
Finally, from what is known on radiative decays in $^{14}$N
\cite{ajz91}, it is difficult to see how any of the observed $\gamma$ rays
could originate from higher levels in \lam{15}{N}. 
In this section, then, we accept the level assignments made in 
Figs.~\ref{oxlevel} and \ref{nilevel} and compare the experimental results 
with theory.

 As outlined in Sec.~\ref{sec:intro-calculation}, the structure of $p$-shell
$\Lambda$-hypernuclei is interpreted in terms of shell-model calculations
that include both $p^ns_\Lambda$ and $p^ns_\Sigma$ configurations. These
calculations, including preliminary analyses of the present data, are
described in Ref.~\cite{millener05}, and in more detail
in Ref.~\cite{millener07}. The $\langle p_Ns_\Lambda|V|p_Ns_\Sigma\rangle$ 
matrix elements were calculated from  a multi-range Gaussian 
potential (YNG interaction) fitted to G-matrix elements calculated
for the SC97f(S) interaction \cite{akaishi00}. Harmonic oscillator 
wave functions with $b=1.7$ fm were used. These matrix elements were 
then multiplied by a factor of 0.9 to simulate the $\Lambda$-$\Sigma$ 
coupling of the SC97e(S) interaction. This interaction is based on 
the NSC97e potential \cite{rijken99} which describes quite well the 
spacing of the $1^+$ and $0^+$ states of the $A\!=\!4$ 
hypernuclei~\cite{akaishi00,hiyama02,nogga02,nemura02}. 
In the same parametrization as for the $\Lambda N$ interaction (see 
Sec.~\ref{sec:intro-calculation}),
\begin{equation}
 \bar{V}' = 1.45\quad \Delta' = 3.04\quad S_\Lambda' = S_N' = -0.09
\quad T' = 0.16
\label{eq:lamsig}
\end{equation}
characterize the five matrix elements (in MeV) for the $\Lambda$-$\Sigma$ 
coupling interaction. This interaction \cite{millener05} is kept fixed 
in the present analysis.

\subsection{The $^\mathbf{16}_\mathbf{~\Lambda}$O spectrum}
\label{sec:discussion-16o}

\begin{table*}[t]
\caption{Energy spacings in \lam{16}{O}. $\Delta E_C$ is the contribution 
of the core level spacing (see Fig.~\ref{oxlevel}). The first line in 
each case gives the 
coefficients of each of the $\Lambda N$ effective interaction parameters
as they enter into the spacing, while the second line gives the actual 
energy contributions to the spacing in keV using the parameters in 
Eq.~(\ref{eq:param16}) which are derived from a  fit to the measured level
spacings in the final column. The calculated spacings are given
in the penultimate column. \label{tab:lo16}}
\begin{ruledtabular}
\begin{tabular}{rrrrrrrrrr}
  $J^\pi_i$ & $J^\pi_f$  & $\Delta E_C$ & $\Lambda\Sigma$ & $\Delta$ 
& $S_\Lambda$ & $S_{N}$ &  $T$ & $\Delta E^{th}$ &  $\Delta E^{exp}$ \\
\hline
 $1^-_1$  & $0^-_1$ &  &  &  $\!-0.370$ & 1.367  &  $\!-0.002$ &  $7.888$  
& & \\
 &  & 0 & $\!-32$   &  $-107$  &   $-21$ &  $1$ &   $177$  &  26  & 26 \\
$1^-_2$ & $1^-_1$  & &  & $\!-0.260$ & $\!-1.235$  &  $\!-1.495$ &  
$\!-0.779$ & \\
 & & 6176 & $\!-38$ &  $-75$  & 19 & $489$ & $-17$  &  6536 & 6536 \\
  $2^-_1$ & $1^-_2$ & & &  $0.629$ & 1.367  &  $\!-0.003$ &  $\!-1.710$ & & \\
 & & 0 & $92$   &  $182$  &   $-21$ &  $1$ &   $-38$  &  226 & 224 \\
\end{tabular}
\end{ruledtabular}
\end{table*}

\begin{table*}
\caption{Energy spacings in \lamb{15}{N}. The core contributions 
$\Delta E_C$ to the energy spacings are derived from the excitation 
energies of the core $0^+;1$ and $1^+;0$ states at 2313 and 3948\,keV (see
Fig.~\ref{nilevel}). The first line in each case gives the coefficients 
of each of the $\Lambda N$ effective interaction parameters as they enter 
into the spacing while the second line gives the actual energy contributions 
to the spacing in keV using the parameters in Eq.~(\ref{eq:param16})
which are derived from a fit to the \lam{16}{O} levels spacings.
The full calculated and measured spacings are given in the final two 
columns. The parenthetic values correspond to spin assignments that are not 
experimentally determined but are the likely assignments.
\label{tab:ln15}}
\begin{ruledtabular}
\begin{tabular}{rrrrrrrrrr}
  $J^\pi_i;T_i$ & $J^\pi_{f};T_f$ & $\Delta E_C$ & $\Lambda\Sigma$ & $\Delta$
& $S_\Lambda$ & $S_{N}$ &  $T$ & $\Delta E^{th}$ & $\Delta E^{exp}$ \\
\hline
 $1/2^+_1;0$ & $3/2^+_1;0$ & & & 0.735 & $\!-2.232$ &
0.022 &  $\!-8.921$ & & \\
 & & 0 & 44 &  213  & 33  & $-7$ & $-200$  &  83  & $< 100$ \\
 $1/2^+;1$ & $3/2^+_1;0$ & & & 0.257 & $\!-0.756$  &  0.015
&  $\!-2.957$ & & \\
  &  & 2313 & $\!-57$ & 75 & $11$ & $-5$ & $-66$ & 2268 & 2268 \\
 $1/2^+_2;0$ & $1/2^+;1$& &  & $\!-0.899$ & $\!-0.104$ & $\!-1.363$
& $0.122$ & & \\
 &  & 1635 & $41$ &  $\!-261$  & 2  & $446$ & 3  &  1851 & (1961) \\
 $3/2^+_2;0$ & $1/2^+;1$& &  & 0.471 & $0.028$ & $\!-1.331$
& $\!-0.239$ & & \\
 &  & 1635 & $106$ &  137  & 0  & $435$ & $-5$  &  2304 & (2442) \\
 $3/2^+_2;0$ & $1/2^+_2;0$ &  & & 1.365 & $0.136$
& 0.032 &  $\!-0.361$ & & \\
 & & 0 & 65 &  396  & $-2$  & $-11$ & $-8$  &  453 & (481) \\
\end{tabular}
\end{ruledtabular}
\end{table*}

 From the shell-model calculations, the contribution from each of the
$\Lambda N$ parameters to a given eigenenergy can be calculated and by
taking differences the coefficient of each parameter entering into
an energy spacing can be deduced. These are given in the first line
of Table~\ref{tab:lo16} for each listed pair of levels in \lam{16}{O}.
It can be seen that the coefficients of the parameters giving the 
two doublet spacings do not deviate much from those given in
Eqs.~(\ref{p12}) and (\ref{p32}) describing the simple $jj$-coupling
limit. This is because the $\Lambda N$ interaction
is too weak to cause large mixing between the $1^-$ weak-coupling basis
states with an unperturbed core separation energy of 6.176 MeV. In fact,
the purity of the dominant basis state is never less than 99.7\% for
the states of interest in  \lam{16}{O} and \lam{15}{N}. That is, the 
sum total of admixed $\Lambda$ and $\Sigma$ configurations is less than
0.3\%. The downward energy shifts caused by $\Lambda$-$\Sigma$ coupling 
are calculated to be 29, 61, 99, and 7 keV for the $0^-$, $1^-_1$, 
$1^-_2$, and $2^-$ states, respectively, and these lead to the 
contributions to the energy spacings listed in the fourth column of 
Table~\ref{tab:lo16}. The sum of the contributions from the $\Lambda N$ 
parameters, the core energy difference, and the $\Lambda$-$\Sigma$ 
coupling gives an accurate, but not perfect (see below) representation 
of the energy-level spacings. 

Assuming that the $\Lambda$-$\Sigma$ coupling and the small value 
of $S_\Lambda\!=\!-0.015$ from \lamb{9}{Be} data [see 
Sec.~\ref{sec:intro-previous} and Eq.~(\ref{eq:param7})] are fixed, 
these expressions can be used to extract values for $\Delta$, $S_N$ and 
$T$. Since $S_N$ affects only core separations and not the doublet spacings,
the two measured doublet spacings in \lam{16}{O} (see Fig.~\ref{oxlevel})
give a pair of simultaneous equations for $\Delta$ and $T$. 
The solution gives $\Delta\!=\!0.312$ MeV and $T\!=\!0.0248$ MeV. 
Here, $\Delta$ is derived from the excited-state doublet spacing
in \lamb{16}{O} because the main contributor of this doublet is $\Delta$.
On the other hand, the most important feature of the 
ground-state doublet splitting is the almost complete cancellation
between substantial contributions from $T$ and $\Delta$
(aided by $\Lambda$-$\Sigma$ coupling.)
There is thus great sensitivity to the value of $T$.   

 Similarly, the difference between the centroid energies of the two 
doublets (or the energy separation between the $1^-$ levels), with a 
small correction for $\Lambda$-$\Sigma$ coupling, gives 
$S_N\!=\!-0.322$ MeV; here, $\bm{l}_N\cdot \bm{s}_N$ simply augments 
the spin-orbit splitting of the hole states of the core nucleus $^{15}$O.

 A re-diagonalization of the energy matrices with new parameters leads
to small shifts in the energy levels and slightly changed values for
the coefficients of the parameters in the expressions that give the
energy-level differences. The parameter set (in MeV) that fits the
measured energy-level spacings in \lam{16}{O} is then given by
\begin{equation}
\Delta= 0.290\quad S_\Lambda =-0.015\quad {S}_{N}
 = -0.327 \quad {T}=0.0224 \; . 
\label{eq:param16}
\end{equation}
The second line for each pair of levels in Table \ref{tab:lo16} and
Table \ref{tab:ln15} gives the actual breakdown of the 
contributions to the energy spacings for this parameter set.
As mentioned above, all parameters except for $S_\Lambda$
are derived only from  the \lam{16}{O} levels.
Nevertheless, $\Delta$ and $S_N$ show smaller but similar values
to those in Eq.~(\ref{eq:param7}) obtained from the \lamb{7}{Li} levels.

 It is to be noted that a 0.15\% admixture of the 
$1^-$ basis state in \lam{16}{O} reduces the binding energy
of  the lower $1^-$ state by 10 keV ($0.0015 \times 6500$) and vice versa
for the upper state. This is the reason why the sum of the individual
energy contributions in Table \ref{tab:lo16} (similarly in
Table \ref{tab:ln15}) do not add up to precisely $\Delta E^{th}$.

\subsection{The $^\mathbf{15}_\mathbf{~\Lambda}$N spectrum}
\label{sec:discussion-15n}

 In addition to the effective $YN$ interaction, the structure of \lam{15}{N}
is sensitive to details of the two-hole, $p$-shell wave functions for 
$^{14}$N. The tensor interaction in the $p$-shell Hamiltonian used for
the \lam{15}{N} calculation was kept fixed during a fit to 90 $p$-shell 
levels and was chosen to ensure cancellation in the Gamow-Teller matrix 
element for $^{14}$C($\beta^-$) decay. The relevant core wave functions 
in LS coupling are (these are $p^{10}$ wave functions in a supermultiplet 
basis so that the phases differ from those for simple two-hole wave 
functions)
 \begin{eqnarray}
| {^{14}{\rm N}(1^+_1;0)}\rangle & =\! & -0.1139\,^3S + 0.2405\,^1P - 
0.9639\,^3D \nonumber \\
| {^{14}{\rm N}(1^+_2;0)}\rangle & =\! & ~~0.9545\, ^3S + 0.2958\, ^1P 
 - 0.0390\, ^3D  \nonumber \\
| {^{14}{\rm N}(0^+;1)}\rangle & =\! & ~~0.7729\,^1S + 0.6346\,^3P \; ,
\label{eq:14gs}
\end{eqnarray}
while in $jj$ coupling the states are 85\% $p_{1/2}^{-2}$, 74\% 
$p_{1/2}^{-1}p_{3/2}^{-1}$, and 93\% $p_{1/2}^{-2}$, respectively.
The Gamow-Teller matrix element is very closely related to 
the $\langle\sigma\tau\rangle$ matrix element for the core $0^+;1\to 1^+;0$
M1 transition in $^{14}$N, which is given by 
\begin{equation}
\langle\sigma\tau\rangle\propto \sqrt{3}a(^1S)\,a(^3S) + a(^1P)\,a(^3P)\, ,
\label{eq:gt}
\end{equation}
with the amplitudes to be taken from Eq.~(\ref{eq:14gs}).

 The contributions to the energy spacings from a shell-model calculation
using the parameters of Eq.~(\ref{eq:param16}) are given in 
Table~\ref{tab:ln15}. Starting with the ground state, the energy shifts 
due to $\Lambda$-$\Sigma$ coupling are calculated to be 
59, 15, 116, 75, and 10 keV.

 The coefficients that enter the ground-state doublet spacing 
show a significant shift away from the $jj$-coupling limit, in which 
the coefficients are $-3/2$ times those in Eq.~(\ref{p12}). 
Specifically, the changes from 0.5  to 0.74 for $\Delta$ and from $-12$ 
to $-9$ for $T$ mean that the higher-spin member of the doublet 
is predicted to be the ground state in contrast to the usual ordering for 
$p$-shell hypernuclei, including \lam{16}{O}.
This spacing was measured to be $E(1/2^+_1) - E(3/2^+_1) > -100$ keV and thus 
the calculated spacing, 83 keV, is consistent with the data.

 The excitation energy of the $1/2^+;1$ state, which gets a significant
contribution from $\Lambda$-$\Sigma$ coupling, is well reproduced.
Note that the contribution of $S_N$ is small and would be zero in the $jj$
limit of $p_{1/2}^{-2}$ for both core states. Viewed from another 
perspective, the coefficient of $S_N$ for the $1/2^+;1$ state depends 
strongly on the matrix element connecting the $^1S$ and $^3P$ components 
of the core wave function in Eq.~(\ref{eq:14gs}). These components
are sensitive to the p-shell interaction used. For example, the Cohen 
and Kurath interactions \cite{ck65} give rise to amplitudes of $\sim 0.53$ 
for the $^3P$ component, a smaller coefficient of $S_N$ in the
hypernuclear calculation, and an energy for the $1/2^+;1$ state that 
is higher by more than 100 keV. 

The level spacings of the members of the excited-state doublet 
($3/2^+_2, 1/2^+_2$) from the $1/2^+;1$ 
state and the doublet spacing itself are calculated to be 
2304, 1851 keV and 453 keV, respectively.
This upper-doublet spacing is dominantly given by $\Delta$ 
because the core state \EL{14}{N}($1^+;0$) is mainly $^3S$ in nature.
This doublet spacing is not experimentally determined but the 
observed 2442 and 1961-keV \ggg \ rays are likely to be
the transitions from the upper doublet members.
Comparing the measured energy difference, $481$ or $-481$ keV,  
with the calculated  value, 453 keV, the
assignments shown in the last column of Table \ref{tab:ln15} 
are strongly indicated.  For this assumption,
the doublet spacing is well reproduced using
the parameter set in Eq. \ref{eq:param16}. 
On the other hand, the excitation energies of both members of the
excited-state doublet are underestimated by about 100 keV.
Here, $S_N$ provides the dominant contribution to the
shift of the centroid of the doublet above the unperturbed core
energy separation of 3948 keV and there is some dependence on the 
shell-model wave functions for the core. For example, the coefficient 
of $S_N$ for the third and fourth entries in Table~\ref{tab:ln15} would 
be $-1.5$ in the $jj$-coupling limit.
 
\subsection{Electromagnetic transitions}
\label{sec:discussion-em}

 The influence of the $\Lambda N$ parameters on the spectra of \lam{16}{O} 
and \lam{15}{N} has been discussed in the preceding two subsections and 
some use of the spectra and transition rates from the shell-model
calculations has already been made in interpreting the results of this 
experiment in Sec.~\ref{sec:results}. The electromagnetic lifetimes
and branching ratios for the observed transitions are largely 
controlled by the M1 matrix elements for the three core transitions in
$^{15}$O and $^{14}$N. For the upper doublets in both hypernuclei, the
transition rates are governed by strong M1 transitions. Consequently,
the lifetimes are short ($\sim$ fs) and the de-excitation $\gamma$-ray
lines are strongly Doppler broadened. For similar reasons, the $\gamma$-ray
branching ratio for the $1^-_2$ level of \lam{16}{O} differs little from
the simple weak-coupling estimate (2:1 in favor of the $0^-$ final state).

 The situation is very different for the decay of the $1/2^+;1$ state
of \lam{15}{N} because (1) the measured lifetime of $1.5\pm0.3 \pm 0.3$ ps 
is much longer than 0.1\,ps for the core transition 
(Sec.~\ref{sec:lifetime15}) and (2) the limit of 9\% for the 
$\gamma$-ray branch to the $1/2^+$ member of the ground-state doublet 
is much less than 33\% in the weak-coupling limit (Sec.~\ref{sec:branch15}).

 The B(M1) values are given by
\begin{equation}
 \textrm{B(M1)} = {3\over 4\pi}\, {2J_f + 1\over 2J_i +1}\, M^2\ \mu_N^2\, ,
\label{eq:m1}
\end{equation}
where $M$ is the reduced matrix element of the M1 operator.
The M1 core transitions of interest, the magnetic moments of $^{14}$N,
$^{15}$N, and $^{15}$O, and many other M1 properties of $p$-shell nuclei
are well described by the present wave functions together
with an effective M1 operator~\cite{millener07} that specifies the six M1
matrix elements in the $p$ shell (equivalently, the isoscalar and isovector
$g$ factors for the operators $\bm{l}$, $\bm{s}$, and $[Y^2,\bm{s}]^1$).
For the $0^+;1\to 1^+_1;0$ and  $0^+;1\to 1^+_2;0$ transitions in
$^{14}$N, the values of $M$ are calculated to be $-0.251$ and 2.957, 
respectively~\cite{millener07}. The matrix elements contain similar 
negative orbital contributions while the spin contributions are 
$\sim 0$ and large and positive, respectively, according to 
Eq.~(\ref{eq:gt}).

 The corresponding matrix elements $M$ for the $1/2^+;1$ $\to$ $1/2^+;0$ and 
$3/2^+;0$ transitions in \lam{15}{N} are the same in the weak-coupling
limit and equal to $-0.251$ [the $2\!:\!1$ branching ratio comes from the
$2J_f + 1$ factor in Eq.~(\ref{eq:m1})]. The shell-model calculation
admixes small $1^+_2\times s_\Lambda$ components into the final-state
wave functions with positive amplitudes of 0.050 and 0.032 for the 
$1/2^+$ and $3/2^+$ final states. This reduces the matrix element for
the $1/2^+;1\to 1/2^+;0$ transition from $-0.25$ to $-0.10$ and the 
$1/2^+;1\to 3/2^+;0$ matrix element by a smaller amount. The net
effect is to increase the lifetime of the $1/2^+;1$ state by about
a factor of five over the weak-coupling limit (to $\tau\!=\!0.48$ ps)
while the branch to the $1/2^+;0$ state is reduced to 18\%. These
effects are in the right direction, but not large enough, to
explain the experimental results.

 With the present wave functions, the dominant contributions to
the off-diagonal matrix elements between the $1^+_1;0\times s_\Lambda$ 
and $1^+_2;0\times s_\Lambda$ basis states are from $S_N$ followed by
contributions from $T$ that are constructive for the $1/2^+$ state
and destructive for the $3/2^+$ state~\cite{millener07}. The larger 
off-diagonal matrix element for the $1/2^+$ case means that the 
$\gamma$-ray branch to the $1/2^+$ state is suppressed relative to the 
$3/2^+$ state (unless the mixing is large enough to change the sign
of the M1 matrix element). If the amplitudes of the $1^+_2;0\times s_\Lambda$
admixtures are scaled up (by a factor of $\sim 1.6$) to reproduce the 
measured lifetime $\tau\!=\! 1.5 \pm 0.3 \pm 0.3$ ps, 
the $\gamma$-ray branch to the $1/2^+$ 
state drops to $\sim 3.5$\% which is consistent with the limit derived in 
Sec.~\ref{sec:branch15}. These results are sensitive to the choice of 
$p$-shell wave functions. It is also worth mentioning that the admixtures 
of $\Sigma$ configurations contribute roughly $+0.01$ to the M1 matrix 
elements in either case~\cite{millener07} and that these are significant 
compared to the $M\!=\!-0.083$ for the $1/2^+;1\to 3/2^+$ transition that 
reproduces the measured lifetime. 

\subsection{Spin-dependent parameters from the present data}
\label{sec:discussion-status}

 Tables \ref{tab:lo16} and \ref{tab:ln15}  show that the parameter
set in Eq.~(\ref{eq:param16}), which was derived to reproduce the
\lam{16}{O} levels while using a fixed $\Lambda$-$\Sigma$ coupling, 
also reproduces quite well the upper-doublet spacing and the 
$1/2^+;1$ excitation energy in \lam{15}{N}, but  predicts a value 
for the excitation energy of the upper doublet in \lam{15}{N} that 
is somewhat too small. The $\Lambda$-spin-orbit parameter $S_\Lambda$ 
is fixed to be small by the $\gamma$-ray data for \lamb{9}{Be} 
\cite{akikawa02,tamura05}.
The theoretical breakdown of the contributions to 
the measured energy spacings in Tables \ref{tab:lo16} and \ref{tab:ln15} 
then shows that $\Delta$, $S_N$, and $T$ are well determined because 
$\Delta$ and $S_N$ dominate in certain spacings while the ground-state 
doublet spacings involve a delicate cancellation between significant
contributions from both $\Delta$ and $T$. It is this cancellation
that fixes $T$ to be small and positive based on the measured spacing
of the ground-state doublet in \lam{16}{O}, in general agreement
with the predictions of $YN$ models, $T\!=\!0.01 - 0.06$ MeV, as 
described in Sec.~\ref{sec:intro-tensor}.

 The matrix elements $\Delta$ and $S_N$ take values that are somewhat 
smaller than the corresponding quantities in Eq.~(\ref{eq:param7}) 
that are determined primarily from \lamb{7}{Li}. This is what would be 
expected if the nuclei at the end of the $p$ shell were significantly
larger than those at the beginning of the shell. However, it is well
known that the charge radii of stable $p$-shell nuclei are almost constant 
throughout the shell \cite{wilkinson66,devries87}, essentially because
the $p$-shell nucleons become more deeply bound for the heavier nuclei.
In fact, if one calculates the $\Lambda N$ matrix elements from a YNG 
interaction using Woods-Saxon wave functions, they tend to be slightly
larger for $A\!=\!16$ than for $A\!=\!7$ \cite{millener01}. This is what
led to the choice of $\Delta\!=\!0.468$ MeV in the first analysis of our
\lam{16}{O} data \cite{ukai04}. Now, the data on the upper doublets
in \lam{16}{O} and \lam{15}{N} (with some reservations) imply a smaller 
value of $\Delta\! =\! 0.290$ MeV.

 Confidence in this value of $\Delta$ is increased by the
fact that the ground-state doublet spacing in \lam{11}{B} has been
established as 264 keV \cite{miura05,hashtam06,ma07} and is reproduced 
by this value of $\Delta$ \cite{millener07}. It is notable that 
a value $\Delta\! < \! 0.30$ MeV was proposed \cite{fetisov91} 
to account for the non-observation of the ground-state doublet 
transition in \lam{10}{B}
above 100 keV in the first hypernuclear experiment with Ge $\gamma$-ray
detectors \cite{chrien90} (a result confirmed in the present experiment
using a $^{10}$B target \cite{hashtam06}).

\section{Summary}
\label{sec:summary}

 A $\gamma$-ray spectroscopy experiment on \lam{16}{O} and \lam{15}{N} was
performed at the BNL-AGS D6 beamline employing a high quality 0.93 GeV/c 
$K^-$ beam and the Hyperball Ge detector array. The experiment is one of
a series aimed at studies of spin-dependent $\Lambda N$ interactions
through the precise measurement of $\gamma$-ray transitions in $p$-shell 
hypernuclei. The bound states of both hypernuclei were produced via the 
\EL{16}{O}\Kpi\ reaction. We succeeded in observing three 
\ggg -ray transitions in \lam{16}{O} and three in \lam{15}{N}.

 For \lam{16}{O}, we determined the level scheme for the four bound 
negative-parity states [two doublets, ($1^-_1$, $0^-$) and 
($2^-$, $1^-_2$)] from these \ggg \ rays, 
although another assignment for the $2^-$ state is not excluded.
In particular, we determined the excitation energy of the 
$1^-_2$ state to be  $6561.7 \pm 1.1 \pm 1.7$ keV and found
a small spacing of $26.4 \pm 1.6 \pm 0.5$ keV for the ground-state 
doublet ($1^-$, $0^-$) with the $0^-$ state being the ground state.
The doublet spacing determines a small but nonzero strength for 
the $\Lambda N$ tensor interaction and this is the first experiment
to give direct information on the $\Lambda N$ tensor interaction.
One of the three  \ggg \ rays  in \lam{16}{O} is likely to be a
transition from the $2^-$ spin-flip state to the one of the ground-state 
doublet members ($2^-\to 1^-_1$) and this constitutes the first 
observation of the direct production of a spin-flip state via the
\Kpi\ reaction. 

 For \lam{15}{N}, we observed a rather sharp 2268-keV $\gamma$ ray
and measured the corresponding lifetime via the Doppler-shift
attenuation method to be $1.5\pm 0.4$ ps. The $\gamma$ ray was
interpreted as a transition from the $1/2^+;1$ level of \lam{15}{N} to
the $3/2^+;0$ member of the ground-state doublet. Because the transition
to the $1/2^+_1;0$ member of the ground-state doublet was not observed
(a limit of $< 9$\% was put on the $\gamma$-ray branch), 
the spacing and the spin ordering of the ground-state doublet
were not determined, but we obtained an upper limit on the spacing energy
of $E(1/2^+_1)-E(3/2^+_1)> -100$ keV. We also observed the \ggg -ray 
transitions which can be assigned to those
from the upper-doublet states $(3/2^+_2, 1/2^+_2)$ to the 
$1/2^+;1$ state.

 We also measured the reaction angle ($\theta_{K\pi}$) distributions of 
the \lam{16}{O} and \lam{15}{N} \ggg \ rays. Analysis of the distributions 
of \lam{15}{N} \ggg \ rays provides information on the spins of the initial 
states of \lam{16}{O} decaying to the excited states of \lam{15}{N} via
proton emission.

The level spectra obtained for \lam{15}{N} and \lam{16}{O} are
consistently explained by the set of values for the $\Lambda N$
interaction parameters $\Delta$, $S_N$, and $T$ in Eq.~(\ref{eq:param16}).
The determination of $T$ was, in fact, the main motivation for the
present experiment.

\begin{acknowledgments}
The authors would like to thank the BNL-AGS staff for support of 
the experiment.
This work is supported by the U.~S.~DOE under Contract No. DE-AC02-98CH10886,
by Grants-in-Aid Nos.~11440070 and 15204014 for Scientific Research from
the Ministry of Education of Japan,
and a Grant-in-Aid No.~1507122 for JSPS Fellows.   
\end{acknowledgments}

\end{document}